\documentclass[10pt]{article}
\usepackage[a4paper,margin=2.5cm]{geometry}
\usepackage{microtype}   

\usepackage{amsmath,amssymb,bm, bbm, graphicx, xcolor}
\usepackage[percent]{overpic}
\usepackage[
    colorlinks=true,
    linkcolor=magenta,
    citecolor=blue,
    urlcolor=black,
]{hyperref}
\usepackage{tikz}
\usepackage{standalone}
\usetikzlibrary{
    decorations.pathreplacing,
    decorations.pathmorphing,
    arrows.meta,
    backgrounds,
    calc
}
\usepackage{etoolbox}
\apptocmd{\appendix}{
  
  \pretocmd{\section}{\setcounter{equation}{0}}{}{}
}{}{}

\usepackage[
  backend=biber,
  style=numeric-comp,
  sorting=none,
  doi=true,
  isbn=false,
  url=true,
  eprint=false
]{biblatex}

\AtEveryBibitem{
  \clearfield{urldate}
  \clearfield{urlyear}
  \clearfield{urlmonth}
  \clearfield{urlday}
}

\newenvironment{acknowledgments}
{
  \section*{Acknowledgments}
  \addcontentsline{toc}{section}{Acknowledgments}
}
{}

\title{\Large\textbf{Balancing structure and randomness: maximum entropy networks for context-dependent computations}}
\author{Ludwig Hruza\thanks{ludwig.hruza@ens.psl.eu} \,and Srdjan Ostojic\thanks{srdjan.ostojic@ens.psl.eu} \\[0.3cm]
\normalsize Laboratoire de Neurosciences Cognitives et Computationnelles \\
\normalsize INSERM U960, École Normale Supérieure - PSL Research University, Paris, France
}
\date{\today
\\[5cm]
\begin{minipage}{0.9\textwidth}
\raggedright
\textbf{Classification:} Biological Sciences/Neuroscience, Physical Sciences/Biophysics and Computational Biology\\[0.3 cm]
\textbf{Keywords:} Neural Networks, Computational Neuroscience, Machine Learning  \\[0.3 cm]
\textbf{Corresponding author:} Srdjan Ostojic, Departement des Études Cognitives de l'ENS, 29 rue d'Ulm 75005 Paris,
+33 (0)1 44 32 26 44, srdjan.ostojic@ens.psl.eu \\[0.3cm]
\textbf{Preprintserver:} https://arxiv.org/abs/2605.25607
\end{minipage}
}

\begin{document}
\maketitle

\newpage

\begin{abstract}
Understanding how network function constrains neural connectivity is a central challenge in neuroscience. An influential approach is to train neural networks with gradient descent on cognitive tasks and characterize the resulting connectivity. A key limitation is that the resulting structure depends on the details of the training procedure. Here we propose a complementary normative approach based on the maximum entropy principle for network connectivity, independent of any particular learning algorithm. We describe connectivity as a probability distribution over single-neuron weights, express task requirements as constraints on this distribution, and determine the unique distribution maximizing Shannon entropy subject to these constraints. A weight scale parameter controls the balance between randomness and task-induced structure. We apply this framework to context-dependent input-selection tasks in 2-layer feed-forward networks, and show that maximum entropy inference becomes analytically tractable by mapping nonlinear networks onto gain-modulated linear models. Starting from an a priori homogeneous distribution, we find that maximizing entropy under task constraints leads to the emergence of populations of neurons, each defined by its pattern of contextual gain modulation. Increasing the number of contexts drives a transition from context-specialized to unspecialized, random populations. Increasing the weight scale drives a parallel transition from structured to random stimulus selectivity. Strikingly, this maximum entropy connectivity matches both qualitatively and quantitatively the structure of networks trained with gradient descent across different learning regimes. Our results suggest that the interplay between task constraints and entropy maximization provides a fundamental principle for understanding the relationship between structure and function in neural networks.
\end{abstract}

{\vspace{3cm}
\renewcommand{\abstractname}{Significance Statement}
\begin{abstract}
A central question in neuroscience is how the brain's synaptic connections are organized to support flexible behavior. A popular approach is to train artificial neural networks on behavioral tasks and study the resulting connectivity, but the emerging structures depend on arbitrary choices in the training, making it hard to identify general principles. We introduce a complementary approach based on the maximum entropy principle: given the constraints imposed by a task, what is the most random connectivity consistent with those constraints? Applying this framework to a context-dependent task, we find that task constraints alone drive the emergence of distinct neuronal populations with interpretable connectivity structure. Remarkably, this principled, training-free prediction quantitatively matches the connectivity of networks trained with machine learning algorithms.
\end{abstract}
}

\newpage
\tableofcontents
\section{Introduction}
One of the  fundamental goals of neuroscience is  to understand how the structure of neural connectivity gives rise to the organization of neural activity and, ultimately, to computations and behavior.
Despite high levels of biological variability and apparent randomness, steady progress in experimental techniques  has been uncovering increasing  levels of statistical structure in measurements of both connectivity \cite{Devineni2024-ht, MICrONS-Consortium2025-we} and activity \cite{Ebitz2021-gy,Chung2021-ao, Kaufman2022-qn,ostojic_ComputationalRole_2024}. Parallel to these experimental efforts, theoretical works have sought  normative principles that determine how the function of a network constrains its structure \cite{Barlow1961-bj,Field1994-qz,Atick1990-db,Fusi2016-aa, Litwin-Kumar2017-bw}. A particularly influential approach has been to  train networks with gradient descent on experimentally relevant cognitive tasks \cite{Sussillo2014-aj, Barak2017-tr, Richards2019-sz, Yang2020-fe, Saxe2021-tl, Yang2021-mc,Kanwisher2023-yv}, and then characterize the structure of the resulting neural dynamics and   connectivity weights \cite{Zipser1988-ez, mante_ContextdependentComputation_2013,  Yang2019-wi, dubreuil_RolePopulation_2022, Johnston2023-cr, driscoll_FlexibleMultitask_2024,Johnston2024-xd}. While such analyses provide a powerful framework for generating hypotheses about the relationship between connectivity, dynamics and function, 
it has been debated to which extent the results depend on the details of the training algorithm and the hyperparameters used for gradient descent \cite{Maheswaranathan2019-tz,Mehrer2020-os,Turner2021-on}. In particular, recent works have shown that the amount of structure in the trained network depends on the learning regime set by the initialization of the parameters \cite{flesch_OrthogonalRepresentations_2022, Schuessler2024-bg, Liu2023-bk}. General principles governing the interplay between structure and randomness in networks of neurons have therefore remained elusive.

Here we introduce a complementary approach to this question based on the maximum entropy principle, which allows us to infer connectivity from task constraints alone, without relying on gradient descent optimization. The number of constraints associated to a cognitive task is typically far smaller than the number of network parameters, making gradient descent optimization a highly underdetermined problem. Rather than sampling, in a potentially biased way, the large space of zero-loss solutions our approach goes as follows: (i) we describe connectivity as a probability distribution of single-neuron weights, which is natural for networks with a wide hidden layer \cite{mei_MeanField_2018, rotskoff_TrainabilityAccuracy_2022,sirignano_MeanField_2019}; (ii) next we deduce a finite number of constraints on the moments of this distribution that ensure compatibility with the task; (iii) finally we optimize the distribution to maximize its Shannon entropy given the constraints \cite{Jaynes1957-wg}. This yields the minimally structured connectivity distribution that is sufficient to perform the task, a unique distribution that is as random as possible given the imposed constraints, and therefore free of potential artifacts introduced by any particular training procedure. We call {\em maximum entropy networks} the models obtained by sampling connectivity weights from this maximum-entropy distribution.

We  apply this approach to a class of context-dependent tasks \cite{Cohen1990-eh,mante_ContextdependentComputation_2013, Barak2013-my, Rodgers2014-pk,Saez2015-ox, mastrogiuseppe_EvolutionNeural_2023,pagan_IndividualVariability_2025, srinath_CoordinatedResponse_2024}, in which several stimulus inputs have to be combined linearly in different ways that depend on a contextual signal. These tasks are a hallmark of flexible behavior \cite{Sakai2008-gh,Okazawa2022-aw}, and are also computationally interesting precisely because they are  non-linear in the joint combination of stimulus and context, yet linear in  the stimulus alone. We focus on the simplest architecture capable of implementing context-dependent input selection:  feed-forward networks with one hidden layer. Linearizing such networks in each context, we map them onto the class of \textit{gain-modulated linear models}, which are closely related to gated linear networks \cite{Saxe2022-li}. In the gain modulated linear model, task constraints take on a simple expression in terms of third-order statistics of network parameters. To control the interplay between structure and randomness, we add a second set of constrains  which  fixes the overall scale  of synaptic weights. With these constraints, we show that 
the maximum entropy inference is mathematically tractable, and leads to connectivity distributions with a particularly interpretable and non-trivial structure. We examine this emerging structure as function of two hyperparameters, the number of contexts and the overall weight scale.

Our central finding is that combining task constraints with entropy maximization leads to the emergence of a population structure in the connectivity distribution that is a priori homogeneous. Each population is defined by its pattern of gain values across the different contexts.
Within each population, the joint distribution of input and output weights is Gaussian, with a covariance structure that determines the selectivity of the neurons to the different inputs, and the manner in which they contribute to the output. The overall populational organization, and the structure of the weight distribution, depend on the number of context and the  scale of synaptic weights which controls the balance between randomness, i.e.\ high entropy, and structure induced by task constraints. Increasing the number of contexts leads to a transition from context-specialized to unspecialized, random populations, while increasing the weight scale induces a parallel transition from structured to random stimulus selectivity. Comparing the maximum entropy distribution with the distribution of single neuron parameters obtained using gradient descent, we find that  the  statistics of trained weights  match both qualitatively and quantitatively with the maximum entropy distribution, across different learning regimes of gradient descent. In particular, we find the same transition from structured to random selectivity with varying initialization scale of the weights \cite{flesch_OrthogonalRepresentations_2022}, so that controlling the balance between structure and randomness, maximum entropy networks interpolate between different types of models for context-dependent computations proposed in earlier works \cite{Cohen1990-eh,mante_ContextdependentComputation_2013}.

\section{Maximum entropy networks}
We first outline our general approach for inferring network parameters without gradient descent.
We consider a one-hidden layer network of the form
\begin{equation} \label{eq:FF-network}
    \hat f(x) = \frac 1 N \sum_{i=1}^N w_i \phi(B_i^T x),
\end{equation}
where $x\in \mathbb R^D$ is the input, $B_i=(B_{i1},\cdots,B_{iD}) \in \mathbb R^D$ is the set of input weights to neuron $i$, $w_i\in\mathbb R$ are the output weights, and $\phi$ is the single-neuron non-linearity. Let us denote the parameters of a single neuron $i$ collectively by $\theta_i=(w_i, B_i)$ and define $h(\theta_i;x):=w_i \phi(B_i^T x)$.
The output  $\hat f(x)$ is a sum of $N$  terms, where each term $h(\theta_i;x)$ depends only on the parameters $\theta_i$ of neuron $i$.
It can therefore be interpreted  as an empirical average  over neurons in the network. When the number of neurons $N$ becomes large, this empirical  average typically converges to an average over a smooth distribution $p (\theta)$  \cite{mei_MeanField_2018, rotskoff_TrainabilityAccuracy_2022,sirignano_MeanField_2019}:
\begin{equation}\label{eq:mean-field-network}
    \hat f(x)  = \mathbb E[h(\theta;x)].
\end{equation}
Here $p(\theta)$ is the distribution of single-neuron parameters $\theta_i$ over the units in the network, and 
the structure of that distribution  determines the output of the network. In the limit of large $N$, also called the mean-field limit, optimizing the network's weights  on some training data is therefore equivalent to finding the optimal distribution $p$ of single-neuron parameters. However, this distribution is typically not fully determined by the training data alone. Since the network is in the over-parametrized regime, there is usually a whole set of distributions $p$ with minimal loss. 

Instead of sampling possible solutions with gradient descent, here examine the distribution that in addition to fitting the training data, also maximizes the Shannon entropy
\begin{equation}
    H(p)= - \int p(\theta) \log p(\theta)\,d\theta.
\end{equation}
This maximum-entropy distribution has the advantage of being unique (Appendix \ref{subapp:recap_convex_optimization}). Moreover,
from the point of view of information theory \cite{Jaynes1957-wg}, it is the most agnostic choice consistent with the training dataset, in the sense that is as random as possible and therefore does not make hidden assumptions about additional constraints.

\section{Context dependent input selection}

\begin{figure}[t]
\centering
\begin{overpic}[width=0.7\linewidth]{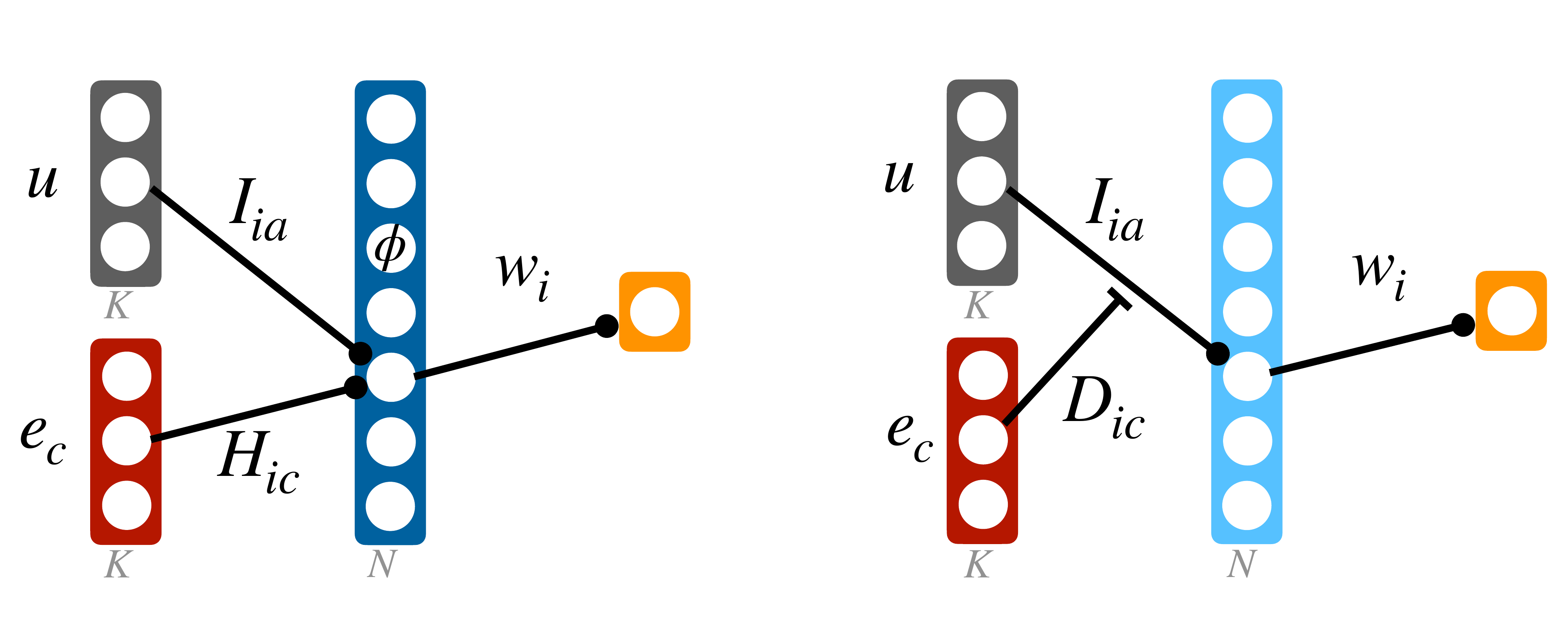}
    \put(-5,35){\textbf{A}}
    \put(50,35){\textbf{B}}
\end{overpic}

\caption{ Model structure. \textbf{A}: We start from a standard feed-forward network with a hidden layer of size $N$, receiving $K$ stimuli $u=(u_a)_{a=1}^K$ and one of $K$ contextual signals $e_c=(\delta_{ac})_{a=1}^K$ through input weights $I,H \in \mathbb R^{N\times K}$ and output weights $w\in \mathbb R^N$, and with a non-linear
activation function $\phi$ (Eq.~\eqref{eq:nonlin_ntwk}). \textbf{B}: We map this model to a  gain-modulated linear network, where each contextual input is replaced by a  gain pattern $D_c=De_c\in \mathbb R^N$ that modulates inputs multiplicatively in an otherwise linear network (Eq.~\eqref{eq:gain-mod-net}).
}
\label{fig:FF-network}
\end{figure}

We apply the maximum entropy approach to a context-dependent input selection task inspired by neuroscience experiments \cite{mante_ContextdependentComputation_2013, siegel_CorticalInformation_2015, pagan_IndividualVariability_2025, srinath_CoordinatedResponse_2024}. We focus on a version
that does not involve temporal integration and can therefore be solved with a feed-forward network. The network receives  a set of $K$ continuous-valued stimuli $u=(u_1,\cdots,u_K)\in \mathbb R^K$ and one out of $K$ contextual signals $c=1,\cdots,K$. It then needs to output the stimulus matching the context,
\begin{equation} \label{eq:task}
    f(u;c)=u_c.
\end{equation}
We represent the contextual signal by a one-hot vector $e_c\in \mathbb R^K$. We denote stimulus input weights to neuron $i$ as $I_i \in \mathbb R^{ K}$, context input weights as $H_i\in \mathbb R^{ K}$ and readout weights as $w_i\in \mathbb R$. 
The output of the network (Fig.~\ref{fig:FF-network} A) is then given by
\begin{equation} \label{eq:nonlin_ntwk}
    \hat f(u,e_c)= \frac 1 N \sum_i w_i\phi(I_i^Tu + H_i^T e_c).
\end{equation}
The choice for the non-linearity $\phi$ is discussed in the following paragraph.

A key property is that this task is linear in the stimulus, but highly non-linear in the combination of stimulus and context. In particular, it can be seen as a $K$-dimensional, continuous version of an XOR computation.

\subsection{Gain-modulated linear networks}
A standard approach for analyzing networks performing context-dependent task is to linearize them within each context \cite{mante_ContextdependentComputation_2013, pagan_IndividualVariability_2025}.
Rather then first training non-linear networks and then linearizing them, here we replace from the outset
our non-linear network by a {\em gain-modulated linear network}, a type of interpretable model we introduce here. We then determine the optimal parameter distribution for this reduced model, and later compare it with the original fully non-linear network.

Linearizing Eq.~\eqref{eq:nonlin_ntwk} around $u=0$ for each context $c$, one obtains
\begin{equation}\label{eq:linearization}
    \hat f(u,e_c)\approx \frac 1 N \sum_i w_i \left(\phi(H_{ic}) + \phi'(H_{ic})I_{i}^T u \right).
\end{equation}
The first term (zero-order) is independent of the stimulus, and therefore does not contribute to the computation. We will assume it is zero for any $c$, and later show that this assumption is correct for the solutions we find (Appendix \ref{subapp:derivation-E[wphi(H)]=0}). The network output in context $c$ is then
\begin{equation}\label{eq:gain-mod-net}
    \hat f(u,e_c)\approx  \frac 1 N \sum_{i,a} w_i D_{ic} I_{ia} u_a.
\end{equation}
where 
\begin{equation}
    D_{ic}:=\phi'(H_{ic})
\end{equation}
acts as a gain on neuron $i$ that modulates in context $c$ the otherwise linear impact of the stimulus  onto its output.
Changing  variables from $H $ to $D$, we replace the contextual input $H_{ic}$ to neuron $i$ in  context $c$  by a gain $D_{ic}$
which  we assume to be positive and bounded by $1$.
We therefore obtain a model that is linear in each context, but non-linear across context. We call this type of model a \textit{gain-modulated linear network} (Figure \ref{fig:FF-network} B). Within this model, each neuron $i$ is characterized by $2K+1$ scalar parameters: $K$ input weights $(I_{i1},\cdots,I_{iK})$, one output weight $w_i$, and $K$ gain parameters $(D_{i1},\cdots,D_{iK})$ that determine the activity of the neuron in each context.

For concreteness, will consider two different non-linearities,
\begin{equation}\label{eq:non-linearity}
   \phi(x)=
   \begin{cases}
       \mathrm{ReLU}(x)\\[2pt]
       \int_0^x e^{-y^2}dy.
   \end{cases}
\end{equation}
In the first case, $\phi'(x)=\Theta(x)$ is a Heaviside function and gains $D_{ic}\in \{0,1\}$ are binary variables. In the other case, $\phi'(x)=e^{-x^2}$ is a Gaussian function and gains $D_{ic}\in [0,1]$ are continuous variables.

\subsection{Mean-field assumption and task constraints.}



In the mean-field limit, all neurons are assumed to be independently and identically distributed. Dropping the neuron index $i$, the output of the gain-modulated linear network in Eq.~\eqref{eq:gain-mod-net} becomes an expectation over the distribution $p(\theta)$ of single-neuron parameters $\theta = (w,I,D)\in \mathbb R^{2K+1}$ with $I=(I_1,\cdots,I_K)$ and $ D=(D_1,\cdots,D_K)$
\begin{equation} \label{eq:output_gain_mod_mf}
    \hat f(u,e_c) = \sum_a \mathbb E[w D_c I_a] u_a.
\end{equation}

Our aim is to infer a probability distribution over the weights of the gain-modulated linear network that is compatible with the task, and maximizes entropy. 
Comparing Eq.~\eqref{eq:output_gain_mod_mf} to Eq.~\eqref{eq:task} one sees that the task constraints can be expressed as
\begin{align} 
    \label{eq:task-constraint}
    \mathbb E[ w D_c I_{a}] &= \delta_{ac}.
\end{align}
In the gain-modulated linear network, the task therefore reduces to $K^2$ constraints on third-order moments of the  single-neuron parameter distribution $p(\theta)$.

On top of these task constraints, we impose the additional set of constraints
\begin{align}\label{eq:scale_constraints}
     \mathbb E[w^2] &=\sigma^2 & \mathbb E[I_a^2]=\sigma^2.
\end{align}
for all $a=1,\cdots,K$. Here $\sigma^2$  is a free parameter that sets the scale of individual weights. Technically these constraints  are required to ensure that the inferred distribution is normalizable, but we will show that they play a key role in controlling the balance between task-constraints and randomness. Here, for simplicity, we assume that the scales of output and input weights are identical, but taking them to be different does not change the results, as the relevant scale is the product of the two (see Appendix \ref{app:maximum_entropy_calculation}).

Above we have described neurons as independent variables right from the start. One could ask what would happen if instead, we had allowed correlations of weights across neurons. In Appendix~\ref{app:iid_neurons} we show that in this case the maximum entropy approach naturally leads to a factorized a distribution of network parameters ${p(\theta_1,\cdots,\theta_N)=\prod_i p(\theta_i)}$. We can therefore concentrate on the distribution of single-neuron parameters without loss of generality.

\section{Maximum Entropy distribution}
To determine the maximum entropy distribution $p$ of single-neuron weights $\theta$ satisfying a given set of constraints $\mathbb E[f_k(\theta)]=c_k$, we minimize a cost function (``Lagrangian'') of the form

\begin{equation}
\mathcal{L}(p,\lambda)=\int p(\theta)\log p(\theta) d\theta -\sum_{k}\lambda_{k}\left(\int f_{k}(\theta) p(\theta) d\theta-c_{k}\right).
\end{equation}
The first term is the (negative) entropy, and the second term consists of Lagrange multipliers enforcing the constraints. Optimizing the entropy term under only the scale constraints Eq.~\eqref{eq:scale_constraints} would lead to a factorized distribution of weights, where $w$ and $I$ would follow a zero-mean Gaussian distribution with variance $\sigma^2$, while $D$ would be uniformly distributed on its support. Adding the task constraints Eq.~\eqref{eq:task-constraint} induces additional structure in the distribution. The task-constraints correspond to third-order correlations between parameters and therefore deviations from a factorized distribution, but we will show that the trade-off between task constraints and entropy also induces additional structure at the level of conditional second-order correlations. Our main goal is to understand this emergent structure of the connectivity distribution.

Importantly, the free parameter $\sigma^2$  controls the balance between task specific structure and entropy. Since $\mathbb E[w D_c I_a]\propto \sigma^2$, a smaller value of $\sigma^2$ leads to more structure, i.e. more alignment between the weights, while a large value of $\sigma^2$ favors entropy and leads to a distribution that almost factorizes.

The maximum entropy distribution depends in principle on $K^2+K+1$ Lagrange multipliers, equal to the number of constraints. However, because of the  symmetry across contexts, they can be reduced to four scalar Lagrange multipliers
 that depend on the weight scale $\sigma^2$ and the number of contexts $K$. These four multipliers obey a set of four non-linear equations (Eq.~\eqref{aeq:abst_eq}) which we solve either numerically or, analytically for large $K$ via a saddlepoint approximation.

Since the constraints Eq.~\eqref{eq:scale_constraints}-\eqref{eq:task-constraint} are only quadratic in the combination of output and input weights $(w,I)$, the maximum entropy  distribution, can be decomposed into a product of a Gaussian and a non-Gaussian part by conditioning on gain parameters $D$ (Appendix~\ref{app:maximum_entropy_calculation}),
\begin{equation}\label{eq:p_wID-decomposed}
    p(w,I,D)=p_D(D)\, p_{wI|D}(w,I|D).
\end{equation}
Here $p_D$ is the marginal distribution of $D$, and is non-Gaussian (Eq.~\eqref{aeq:p_D}).
The remaining part, the joint distribution $p_{wI|D}$ of $(w,I)$ conditioned on $D$, is a zero-mean, $K+1$-dimensional Gaussian with a covariance matrix $\Sigma(D)$ that depends on the gain parameters $D$ (Eq.~\eqref{aeq:Sigma(D)})
\begin{equation}\label{eq:p_wI_gaussian}
    p_{wI|D}=\mathcal N(0,\Sigma(D)).
\end{equation}
As a reminder, $D$ is a vector of $K$ gain values that determine the contribution of each neuron to the output in each of the $K$ contexts. Conditioning on $D$ therefore amounts to defining a population of neurons according to their activity across contexts. The fraction of neurons belonging to  this population in the network is quantified by $p_D(D)$. For each population,
the joint distribution of input and output weights is a multivariate Gaussian with a correlation matrix $\Sigma(D)$.
Our main result is that these distributions are non-isotropic, i.e. the correlation matrices are in general not proportional to the identity.
 Their shapes depend on the gain configuration $D$, as well as on  parameters $\sigma^2$ and $K$. These conditional correlations $\Sigma(D)$ are what we refer to as the emergent structure of the connectivity distribution.

More specifically, the entries in $\Sigma(D)$ are of two types: (i) covariances $\mathbb E[wI_a |D]$ between output weights and input weights corresponding to various stimuli, which determine the output of the network in different contexts; (ii) covariances $\mathbb E[I_a I_b |D]$ between, and variances of, input weights $I_a$ and $I_b$, which determine the selectivity of the population to different stimuli.
Indeed, for the gain-modulated linear network, the points in the $(I_1,I_2)$ plane can  be interpreted as the regression coefficients for how much a given neuron (a dot in the scatter plot) is affected by either of the two stimuli. The corresponding space of  regression coefficients has also been called \textit{selectivity space}, and the structure of the distribution in that space has been used to characterize selectivity \cite{mante_ContextdependentComputation_2013,raposo_CategoryfreeNeural_2014, dubreuil_RolePopulation_2022, ostojic_ComputationalRole_2024}. 
In the following, we therefore denote as {\em random mixed selectivity} the limit where  $(I_a,I_b)$ are distributed as an isotropic Gaussian \cite{raposo_CategoryfreeNeural_2014,ostojic_ComputationalRole_2024}, i.e. the variances are equal and the correlation zero. Conversely, {\em preferential selectivity} to stimulus $a$ corresponds to the limit where the variance of $I_a$ strongly dominates over the variance of any other $I_b$ for the given population. To quantify the selectivity between these two extremes we compute the ratio
\begin{equation} \label{eq:selectivity_index}
    \mathrm{Selectivity}(I_a|D) : = \frac{\mathrm{Var}(I_a|D)-\overline{\mathrm{Var}(I_{b\neq a}|D)}} {\mathrm{Var}(I_a|D)+ \overline{\mathrm{Var}(I_{b\neq a}|D)} }
\end{equation}
where $\overline{\mathrm{Var}(I_{b\neq a}|D)}$ is the average variance of the other input weights conditioned on a configuration of gains $D$.

We next examine how the emergent structure depends on the scale parameter $\sigma^2$, the number of contexts $K$ and  the fact that the gains are binary or continuous.

 \subsection{Binary gains.}
We first focus on the case where the gains are binary, so that each neuron participates in the output in context~$c$ if $D_c=1$ and is silent otherwise. This case directly corresponds to a threshold-linear transfer function (Eq.~\eqref{eq:non-linearity}).

\begin{figure*}[t]
\begin{minipage}[t]{0.49\textwidth}
    \centering
    \begin{overpic}[width=\textwidth]{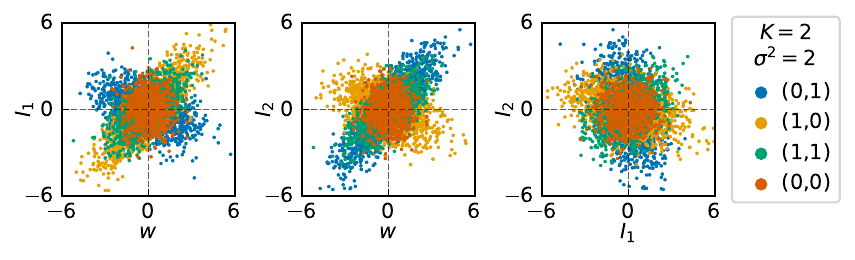}
        \put(0,30){\textbf{A}}
    \end{overpic}
\end{minipage}
\hfill
\begin{minipage}[t]{0.49\textwidth}
    \centering
    \begin{overpic}[width=\textwidth]{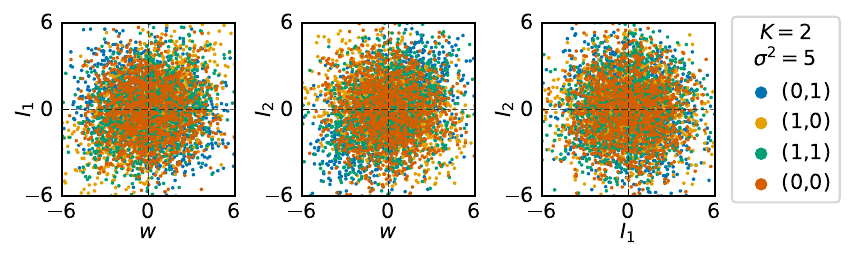}
        \put(0,30){\textbf{B}}
    \end{overpic}
\end{minipage}
\begin{minipage}[t]{0.49\textwidth}
    \centering
    \begin{overpic}[width=\textwidth]{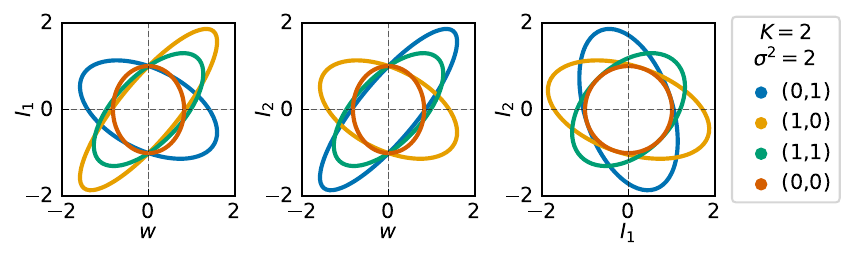}
        \put(0,30){\textbf{C}}
    \end{overpic}
\end{minipage}
\hfill
\begin{minipage}[t]{0.49\textwidth}
    \centering
    \begin{overpic}[width=\textwidth]{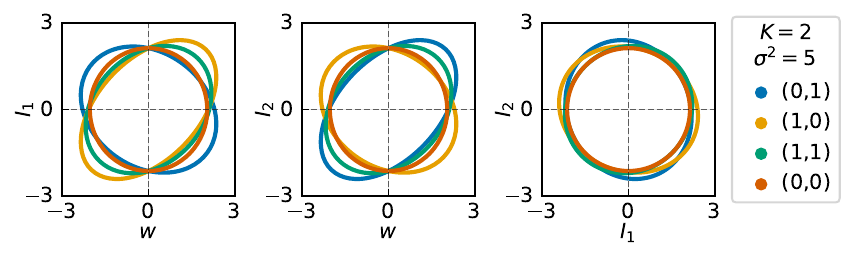}
        \put(0,30){\textbf{D}}
    \end{overpic}
\end{minipage}
\begin{minipage}[t]{\textwidth}
    \centering
    \begin{overpic}[width=.8\textwidth]{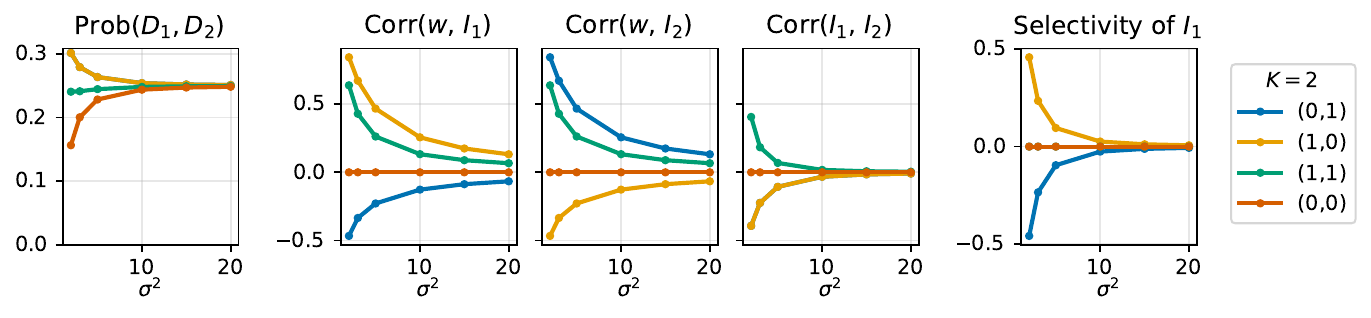}
        \put(0,21){\textbf{E}}
        \put(22,21){\textbf{F}}
        \put(70,21){\textbf{G}}
    \end{overpic}
\end{minipage}
\caption{Maximum Entropy distribution for $K=2$ contexts and binary gains. The  four  configurations of gains  for two contexts define four populations of neurons with $D=(D_1,D_2)=(0,1),(1,0),(1,1)$ and $(0,0)$, represented in different colors. \textbf{A, B}: Samples ($N=5000$) from the maximum entropy distribution for $\sigma^2=2$ (panel A) and $\sigma^2=5$ (panel B), projected onto the planes $(w,I_1)$, $(w,I_2)$ and $(I_1,I_2)$. 
\textbf{C, D}: The covariance matrix of each pair of weights is represented as an ellipse for each population (direction: largest eigenvector, width and height: eigenvalues), for $\sigma^2=2$ (panel C) and $\sigma^2=5$ (panel D) \textbf{E}: Fraction of neurons in each population 
 as a function of the weight scale $\sigma^2\in[2,20]$. \textbf{F}: Correlations between pairs of weights as a function of the scale $\sigma^2$. Here $\mathrm{Corr}(X,Y)=\mathrm{Cov}(X,Y)/(\sigma_X \sigma_Y)$. \textbf{G}: Selectivity of each population to stimulus $a$, measured as the ratio between the difference and sum of $\mathrm{Var}(I_1|D)$ and $\mathrm{Var}(I_{2}|D)$ (Eq.~\eqref{eq:selectivity_index}).
All the quantities were computed from the decomposition Eq.~\eqref{eq:p_wID-decomposed} with covariance matrix $\Sigma(D)$ from Eq.~\eqref{aeq:Sigma(D)}.
}
\label{fig:binary_K=2}
\end{figure*}

\subsubsection{Two contexts}
For $K=2$, the network consists of four populations based on different configurations of gain values $D=(D_1,D_2)$ in the two contexts : $D=(0,0)$.  neurons inactive in both contexts; $D=(0,1),(1,0)$,  neurons active only in one, but not the other context; $D=(1,1)$, neurons active in both contexts. The full distribution 
therefore clusters into four Gaussian populations, with their respective fractions given by the probabilities $p_D(D)$ (Fig.~\ref{fig:binary_K=2}).

In this case, the role of the correlations between input and output weights is particularly transparent. We  illustrate how they contribute to the task computation in context $c=1$, the situation being fully symmetric in context $2$. In context $1$, only populations $(1,0)$ and $(1,1)$ participate in the output. From Eq.~\eqref{eq:output_gain_mod_mf}, the contribution of the stimulus $a=1$ to the output is
\begin{equation}\label{eq:output_relevant_K2_binary}
 \mathbb E[wI_1D_1]= p_D(1,0) \,\mathbb E[wI_1 |(1,0)] + p_D(1,1) \,\mathbb E[wI_1 | (1,1)],
\end{equation}
i.e.\ it is determined by the correlations between $w$ and $I_1$ in the two populations, weighted by the corresponding fraction of neurons. Since $I_1$ is the relevant stimulus in context $1$, according to the task constraints Eq.~\eqref{eq:task-constraint} the two terms need to sum to unity.

The contribution of the stimulus $2$ in context $1$ is similarly determined by the correlations between $w$ and $I_2$ in the two populations,
\begin{equation}\label{eq:output_irrelevant_K2_binary}
 \mathbb E[wI_2D_1]= p_D(1,0) \,\mathbb E[wI_2 |(1,0)] + p_D(1,1) \,\mathbb E[wI_2 | (1,1)].
\end{equation}
Since stimulus two is irrelevant in context 1, according to the task constraints, these two terms need to sum to zero. 

We determine the correlation matrix $\Sigma(D)$ for each population by solving the equations for the four Lagrange multipliers, which reduce to a single equation that we solve numerically (Appendix \ref{subapp:K=2}). Our results show two qualitatively different types of behavior at small and large~$\sigma^2$.

For small values of the weight scale $\sigma^2$ (Fig.~\ref{fig:binary_K=2} A,C), the populations $(0,1)$ and $(1,0)$ contain a greater fraction of neurons than the other two (Fig.~\ref{fig:binary_K=2} E). Neurons therefore tend to specialize to one of the two contexts, although about a quarter of them is active in both ($p_D(1,1)\approx 1/4$). The distribution of weights in the two specialized populations shows a clear difference in correlation structure (Fig.~\ref{fig:binary_K=2} A,C,F): There is a positive correlation between the output and the relevant stimulus ($\mathbb E[wI_1 |(1,0)]>0$ and $\mathbb E[wI_2 |(0,1)]>0$) and a negative correlation with the irrelevant one ($\mathbb E[wI_2 |(1,0)]<0$ and $\mathbb E[wI_1 |(0,1)]<0$). This negative correlation balances the input-output correlation in the population $(1,1)$, which is positive for all stimuli ($\mathbb E[wI_a |(1,1)]>0$, $a=1,2$), ensuring that the contribution of the irrelevant stimulus to the output vanishes (Eq.~\eqref{eq:output_irrelevant_K2_binary}). The two specialized populations show a selectivity preference for the relevant stimulus, as seen in the alignment of the joint distribution of $(I_1,I_2)$ along the relevant stimulus axis (Fig~\ref{fig:binary_K=2} A,C right panels). The context-specialized populations therefore show structured selectivity, while  the neurons in the population $(1,1)$ exhibit random mixed selectivity (Fig~\ref{fig:binary_K=2} G).
 
For large values of the weight scale $\sigma^2$, the neurons in the network are instead equally distributed among the four populations, $p_D(0,1)\approx p_D(1,0)\approx p_D(1,1) \approx p_D(0,0) \approx 1/4$, so that they do not specialize to individual contexts. In particular the fraction of neurons in the population $(0,0)$, which does not participate in the task, becomes equivalent to the other populations. This is the result of the entropy term in the cost function, which favors a uniform distribution of gains  across contexts. For each population, the distribution of synaptic weights becomes increasingly isotropic (Fig.~\ref{fig:binary_K=2} B,D), although the structure of correlations between input and output weights is preserved to satisfy the constraints in Eqs.~\eqref{eq:output_relevant_K2_binary} and \eqref{eq:output_irrelevant_K2_binary} (Fig.~\ref{fig:binary_K=2} F). In the limit of large $\sigma^2$, all populations respond in a similar way to all stimuli and therefore show random mixed selectivity (Fig.~\ref{fig:binary_K=2} G).

\subsubsection{Large number of contexts}

\begin{figure*}[t]
\begin{minipage}[t]{0.49\textwidth}
    \centering
    \begin{overpic}[width=\textwidth]{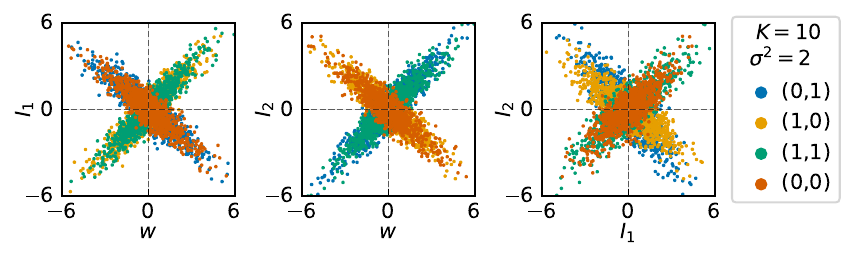}
        \put(0,30){\textbf{A}}
    \end{overpic}
\end{minipage}
\hfill
\begin{minipage}[t]{0.49\textwidth}
    \centering
    \begin{overpic}[width=\textwidth]{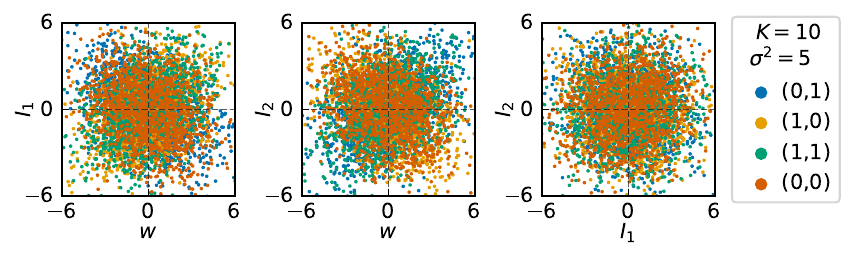}
        \put(0,30){\textbf{B}}
    \end{overpic}
\end{minipage}
\begin{minipage}[t]{0.49\textwidth}
    \centering
    \begin{overpic}[width=\textwidth]{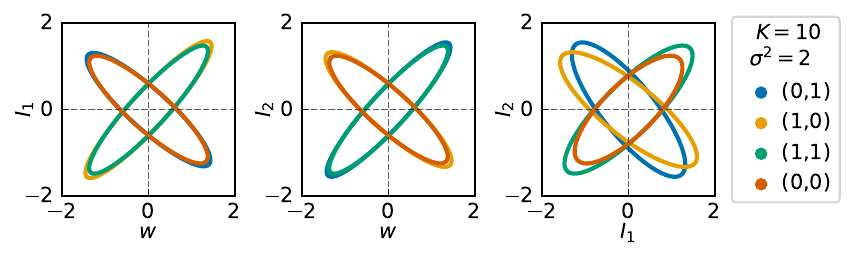}
        \put(0,30){\textbf{C}}
    \end{overpic}
\end{minipage}
\hfill
\begin{minipage}[t]{0.49\textwidth}
    \centering
    \begin{overpic}[width=\textwidth]{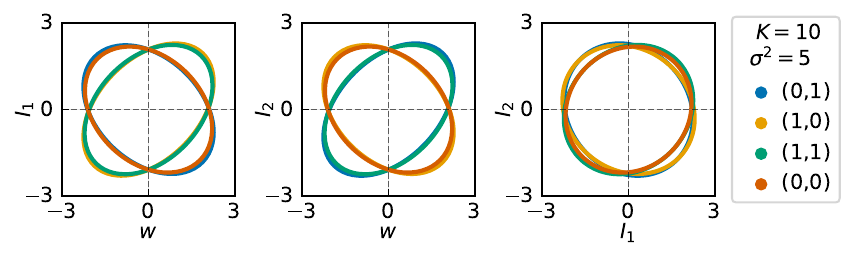}
        \put(0,30){\textbf{D}}
    \end{overpic}
\end{minipage}
\begin{minipage}[t]{\textwidth}
    \centering
    \begin{overpic}[width=.8\textwidth]{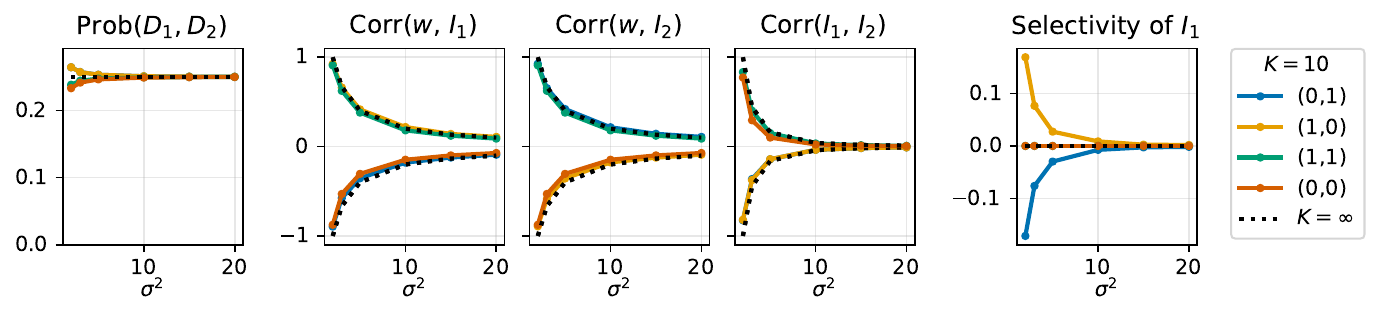}
        \put(0,21){\textbf{E}}
        \put(22,21){\textbf{F}}
        \put(70,21){\textbf{G}}
    \end{overpic}
\end{minipage}
\caption{Maximum Entropy distribution for $K=10$ contexts and binary gains. 
We condition the gain values of the first two contexts $(D_1,D_2)=(0,1),(1,0),(1,1)$ and $(0,0)$,  and average over gain values in other contexts. The four resulting populations are shown in four different colors.  \textbf{A, B}: Samples ($N=5000$) from the maximum entropy distribution for $\sigma^2=2$ (panel A) and $\sigma^2=5$ (panel B), projected onto the planes $(w,I_1)$, $(w,I_2)$ and $(I_1,I_2)$. \textbf{C, D}: The covariance matrix of each pair of weights is represented as an ellipse for each population (direction: largest eigenvector, width and height: eigenvalues)  for $\sigma^2=2$ (panel C) and $\sigma^2=5$ (panel D). \textbf{E}: Probability of the four gain configurations as a function of the weight scale $\sigma^2$. \textbf{F}:~Correlations between pairs of weights as a function of the scale $\sigma^2$. Here $\mathrm{Corr}(X,Y)=\mathrm{Cov}(X,Y)/(\sigma_X \sigma_Y)$. \textbf{G}: Selectivity of each population to stimulus $a$, measured as the ratio between the difference and sum of $\mathrm{Var}(I_1|D)$ and the average of all other variances $I_{a\neq 1}$. All the quantities were computed from the decomposition Eq.~\eqref{eq:p_wID-decomposed} with numerically determined Lagrange multipliers and covariance matrix $\Sigma(D)$ for $K=10$ from Eq.~\eqref{aeq:Sigma(D)}. Black dotted lines represent the asymptotic values $K\to\infty$ obtained from Eq.\eqref{eq:Sigma_large_D}}
\label{fig:binary_K=10}
\end{figure*}

We next examine the situation where the number of contexts $K$ is large. In that limit, the equations for the Lagrange multipliers can be solved analytically (Appendix \ref{subsapp:large-K-limit}), and the resulting distribution has a simple structure. Specifically, the distribution of gain values $p_D(D)$ concentrates on configurations $D=(D_1,\ldots, D_K)$ in which each neuron has an equal number of zeros and ones, randomly distributed across contexts. Conversely, a random half of neurons is active in each context. In contrast to $K=2$,  for $K\to\infty$ individual neurons therefore do not specialize for individual contexts, even for small weight scale $\sigma^2$ which favors task structure over entropy. A direct consequence is that the gain of a neuron in a given context, say $D_1$, becomes independent of the  gains $(D_2,\cdots,D_K)$ in other contexts, and is equal to zero or one with  probability $1/2$.

For such a gain pattern with equal numbers of zeros and ones, the covariance matrix $\Sigma(D)$ takes a simple form  
\begin{align} \label{eq:Sigma_large_D}
\Sigma_{ww} &= \sigma^2  & \Sigma_{wI_a} &= 4 (D_a-1/2) \\ \nonumber
\Sigma_{I_a I_a} &= \sigma^2 & \Sigma_{I_a I_b} &= \frac{16}{\sigma^2}(D_a-1/2)(D_b-1/2),
\end{align}
where we assumed $a\neq b$.
In particular, the covariances between $(w,I_a,I_b)$ only depend on the corresponding gain values $(D_a,D_b)$, but not on the other gain values. Therefore, for fixed $(a,b)$, the entire distribution can be represented by conditioning on $(D_a,D_b)$  (Fig.~\ref{fig:binary_K=10}).

Independently of the value of $\sigma^2$ and $K$, neurons active in a given context $c$ exhibit a positive covariance between the output weights and the weights $I_c$ of the relevant stimulus, i.e. $ \mathbb E[wI_c |D_c=1]\approx 2$ (which is exact for large $K\to\infty$). Conversely, neurons inactive in context $c$ ($D_c=0$) have a negative covariance between output weights and the weights $I_a$ of any irrelevant stimulus, i.e. $\mathbb E[wI_a |D_c=0]\approx -2$ for $a\neq c$. In contrast to the case of two contexts, these covariances are independent of the gains of neurons in other contexts, e.g.  $ \mathbb E[wI_1|(1,0)]=\mathbb E[wI_1|(1,1)]$. For small $\sigma^2$, this structure leads to clear clusters in the weight distributions (Fig.~\ref{fig:binary_K=10} A,C), while increasing $\sigma^2$ leads to increasingly isotropic distributions (Fig.~\ref{fig:binary_K=10} B,D). According to the asymptotic expression in Eq.~\eqref{eq:Sigma_large_D}, the correlation $\mathrm{Corr}(w,I_a|D)=\Sigma_{wI_a}/\sqrt{\Sigma_{ww}\Sigma_{I_aI_a}}$ decays with $1/\sigma^2$. This matches with what we see in Fig.~\ref{fig:binary_K=10} F, already for $K=10$.

This structure has a straightforward interpretation in terms of task constraints. In context $c$, the contribution of the relevant stimulus $c$ to the output should be
\begin{equation}\label{eq:relevant_contribution_largeK}
 \mathbb E[wI_cD_c]= p_{D_c}(1) \mathbb E[wI_c |D_c=1] \overset{\mathrm{task}}{=}1.
\end{equation}
Since $p_{D_c}(1)=1/2$, this implies $\mathbb E[wI_c |D_c=1]=2$ in accordance with Eq.~\eqref{eq:Sigma_large_D}, i.e.\ the positive correlation between the output weights and the weights of the relevant input ensures that the relevant input is selected.
The contribution of any irrelevant stimulus $a \neq c$ instead should be
\begin{eqnarray} \label{eq:irrelevant_contribution_largeK}
    \mathbb E[wI_aD_c] &=&  p_{D_c}(1) \mathbb E[wI_a |D_c=1] \\ \nonumber
    &=& \frac{1}{2} \mathbb E[wI_a] \\ \nonumber
    &=& \frac{1}{2}(\mathbb E[wI_a |D_a=1]+\mathbb E[wI_a |D_a=0]) \overset{\mathrm{task}}{=}0,
\end{eqnarray}
where we used the fact that the covariance of $(w,I_a)$ only depends on $D_a$ (Eq.~\eqref{eq:Sigma_large_D}). This implies \mbox{$\mathbb E[wI_a |D_a=0]=-\mathbb E[wI_a |D_a=1]=-2$}. So once we make the assumption that $(w,I_a)$ is independent from $D_{b\neq a}$, the task constraint is sufficient to determine the entries $\Sigma_{wI_a}$ of the maximum entropy distribution. In other words, for large $K$, the contribution of entropy maximization is to enforce independence of input and output weights from gains values belonging to other context, and to fix the covariance of $(I_a,I_b)$.

The covariance between input weights $(I_a, I_b)$ are positive if $D_a=D_b$, and negative if $D_a\neq D_b$, and proportional to $1/\sigma^2$ (Eq.~\eqref{eq:Sigma_large_D}). The correlation of $(I_a,I_b)$ therefore decays with $1/\sigma^4$ (Fig.~\ref{fig:binary_K=10} F). For small $\sigma^2$, the neurons are therefore organized in two clusters that are selective respectively to the sum and difference of stimuli $a$ and $b$ (Fig.~\ref{fig:binary_K=10} A and C). This  implies strong deviations from random mixed selectivity, even if individual neurons do not specialize for individual contexts, and do not show preferential selectivity to individual stimuli. For larger $\sigma^2$, the joint distributions of $(I_a, I_b)$ become increasingly isotropic (Fig.~\ref{fig:binary_K=10} B and D) implying increasingly random mixed selectivity.
 
\subsection{Comparing continuous and binary gains.}
\begin{figure*}
    \centering
    \begin{overpic}[width = \linewidth]{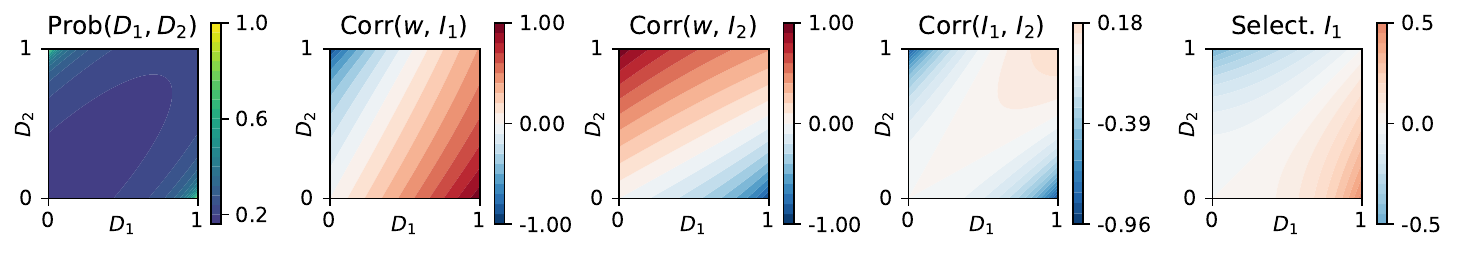}
    \put(0,16){\textbf{A}}
    \put(20,16){\textbf{B}}
    \put(80,16){\textbf{C}}
    \end{overpic}
    \\[0.3cm]
    \begin{overpic}[width = \linewidth]{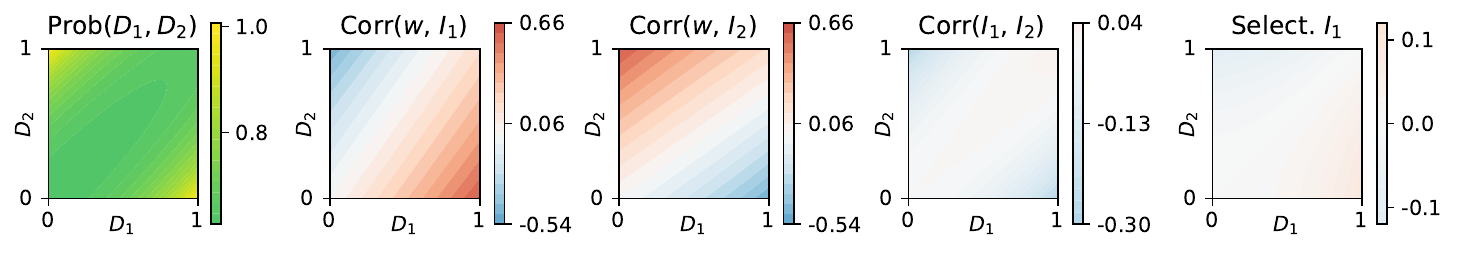}
    \put(0,16){\textbf{D}}
    \put(20,16){\textbf{E}}
    \put(80,16){\textbf{F}}
    \end{overpic}
    \caption{Maximum entropy distribution for continuous gains and $K=2$ as a function of continuous valued $(D_1,D_2)\in[0,1]^2$. \textbf{A, B, C}: Probability, correlation and selectivity for $\sigma^2=4$. The  most probable gain configurations are $(D_1,D_2)=(1,0)$ and $(0,1)$. For these two configurations, the correlations between $w$ and $I_1$ is maximal (equal to $\pm1$) and of opposite sign (panel B, left). For these configurations, there is also preferential selectivity in the input weights since the correlation between $I_1$ and $I_2$ is maximally negative (approximately $-1$, B left), which is reflected in non-zero and opposite selectivity (panel C). \textbf{D,E,F}: Probability, correlation and selectivity for $\sigma^2=10$. Here all gain configuration are approximately equally likely. Beyond this, the correlation structure is similar to the $\sigma^2=4$ case, but weaker. In particular, $I_1$ and $I_2$ are  almost uncorrelated (panel E, right) leading to almost zero selectivity (panel F).}
    \label{fig:continuous_K=2}
\end{figure*}

\begin{figure*}[t]
\centering
\begin{overpic}[width=\linewidth]{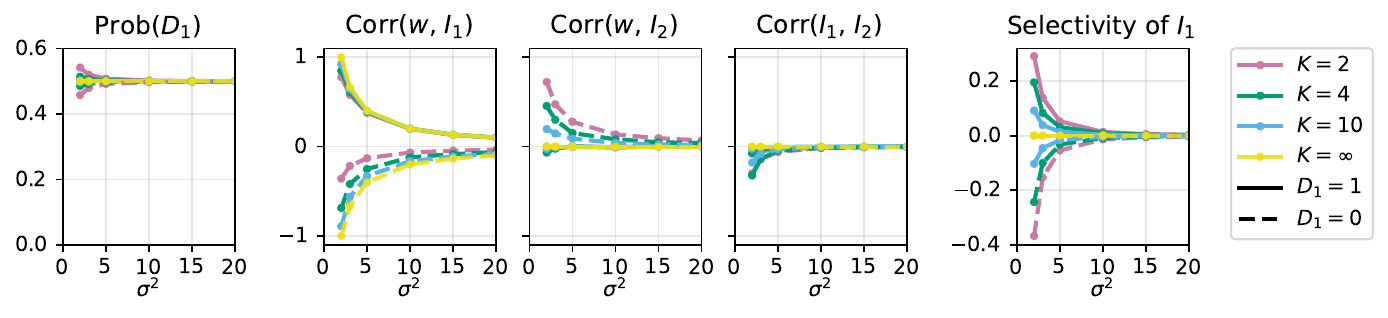}
    \put(0,21){\textbf{A}}
    \put(22,21){\textbf{B}}
    \put(69,21){\textbf{C}}
\end{overpic}
\begin{overpic}[width=\linewidth]{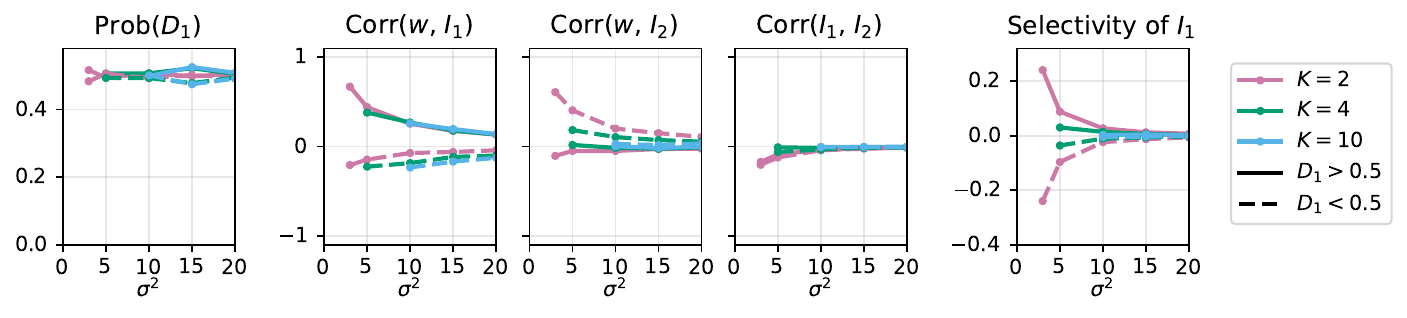}
    \put(0,21){\textbf{D}}
    \put(22,21){\textbf{E}}
    \put(69,21){\textbf{F}}
\end{overpic}
\caption{Comparison between binary (top) and continuous gains (bottom) for different values of $K$. Instead of conditioning on $(D_1,D_2)$ as before, we only condition on the values $D_1$, which corresponds to how the network would operate in context $c=1$. \textbf{A,B,C} Probability, correlation and selectivity for binary gains as a function of the weight scale $\sigma^2$ when conditioned on $D_1=1$ (solid lines) or $D_1=0$ (dashed lines). We also added the asymptotic curves for $K\to\infty$. \textbf{D,E,F}: Same but for continuous gains, conditioned on $D_1>0.5$ (solid lines) or $D_1<0.5$ (dashed lines). For continuous gains there is a $K$-dependent threshold on the minimal value of $\sigma^2$ that is allowed to avoid complex values in the probability distribution (see Eq.\eqref{aeq:constraint_c_Q<1}). We therefore start the curves corresponding to higher values of $K$ at larger $\sigma^2$. Taking into account this difference, the two cases, binary and continuous gains, behave almost identically, even quantitatively.
}
\label{fig:compare_K}
\end{figure*}
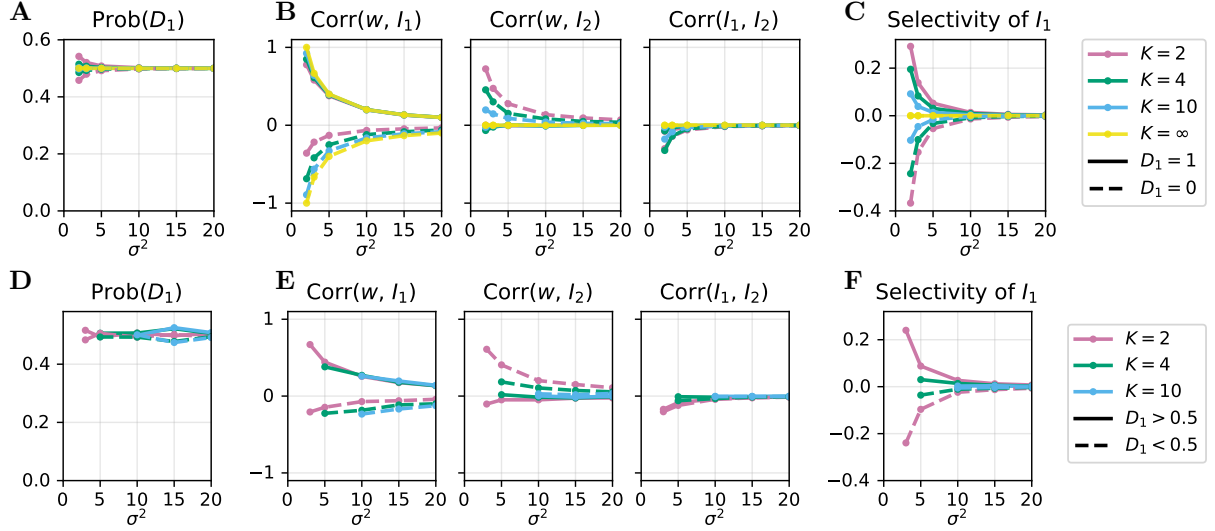

So far we focused on  binary gains, as in that case the neurons in the network can be split in $2^K$ discrete populations with a Gaussian distribution of input and output weights within each of them (Eq.~\eqref{eq:p_wID-decomposed}). 
If the gains are instead continuously distributed in $[0,1]$, as would be the case for a sigmoidal non-linearity (Eq.~\eqref{eq:non-linearity}), the full distribution of network parameters is not anymore a discrete mixture of Gaussians but a continuous one.
The general picture is however preserved if one splits the neurons into subsets depending on whether their gain is smaller or larger than $1/2$. Here we provide an illustration of weight distributions with continuous gains for $K=2$ (Fig.~\ref{fig:continuous_K=2}), and an overall comparison of the binary and continuous cases across different values of the number of contexts $K$ and the weight scale~$\sigma^2$ (Fig.~\ref{fig:compare_K}).

For $K=2$, based on Eq.\eqref{eq:p_wID-decomposed} and Eq.~\eqref{eq:p_wI_gaussian}, the maximum-entropy distribution of network parameters can be fully described by representing as functions of $(D_1,D_2)$ the probability $p_D$ and the pairwise correlations between the output and input weights $(w,I_1,I_2)$ (Fig.~\ref{fig:continuous_K=2}). The resulting picture shows the same qualitative features as in the binary case. For small $\sigma^2$, neurons tend to specialize for one of the two contexts (ie. $p_D$ has maxima  at $(D_1,D_2) = (1,0)$ and $(0,1)$, Fig.~\ref{fig:continuous_K=2}A). Neurons specialized for context $1$ (i.e. neurons with $D_1>1/2$ and $D_2<1/2$) have a positive correlation for $(w,I_1)$ and negative for $(w,I_2)$ (Fig.~\ref{fig:continuous_K=2}B), as well as preferred selectivity for $I_1$ (Fig.~\ref{fig:continuous_K=2}C). The situation is symmetric for neurons specialized for context $2$. As the weight scale $\sigma^2$ is increased, the structure of correlations between input and output weights is preserved, but the gains become more uniformly distributed (Fig.~\ref{fig:continuous_K=2}D) so that the specialization becomes less pronounced (Fig.~\ref{fig:continuous_K=2}E). Moreover, neurons become increasingly mixed selective (Fig.~\ref{fig:continuous_K=2}F).

To compare more systematically the cases of discrete and continuous gains for different values of $K$, we compute
the correlation between pairs of input and output weights, $(w,I_1),(w,I_2)$ and $(I_1,I_2)$, conditioned on the gain $D_1$ (Fig.~\ref{fig:compare_K}). Specifically, we condition on $D_1=1$ or $0$ in the binary case, and $D_1>0.5$ or $D_1<0.5$ in the continuous case. This corresponds to how the network would operate in context $c=1$. The correlation of active neurons (solid lines) between context-relevant input weights $I_1$ and output weights $w$ is positive and decays with increasing weight scale $\sigma^2$ (Fig.~\ref{fig:compare_K}B,E). Notably, this occurs independently of $K$, which shows that this aspect of the structure is preserved for any $K$ and only depends on $\sigma^2$. However, for deactivated neurons (dashed lines), which do not contribute to the task in context $c=1$, the picture differs between small and large $K$ (Fig.~\ref{fig:compare_K}B,E). These neurons are as random as possible and only constrained by the fact that the correlation of $(w, I_1)$ should be zero in any other context $c\neq 1$. 
For large $K$, the distribution of $(w,I_1)$ is independent of any gain $D_{a\neq1}$. Therefore  $\mathbb E[wI_1 |D_1>0.5]$ and $\mathbb E[wI_1 |D_1<0.5]$ must sum to zero, as in the discrete case (Eq.~\eqref{eq:irrelevant_contribution_largeK}), to balance the contribution of the irrelevant stimulus. 
For  $K=2$  and small $\sigma^2$, however, $(w,I_1)$ depends on both $D_1$ and $D_2$ such that $\mathbb E[wI_1 |D_1>0.5]$ and $\mathbb E[wI_1 |D_1<0.5]$ do not need to precisely balance anymore to cancel the contribution of the irrelevant stimulus. Moreover, deactivated neurons in context $1$ are preferably active in context $2$, implying $\mathbb E[wI_2 |D_1<0.5]>0$, while for $K>2$ this correlation vanishes (Fig.~\ref{fig:compare_K}B,E). 

To quantify the selectivity to different stimuli, we compare the variance of the context-relevant input weight $\mathrm{Var}(I_1|D_1)$ to the variance of the other input weights $\mathrm{Var}(I_{a\neq 1}|D_1)$ (Fig.~\ref{fig:compare_K}C,F). For active neurons, with increasing scale $\sigma^2$, we observe a transition from structured to and random mixed selectivity, for both binary and continuous gains.

Similarly to the covariance matrix for binary gains in Eq.~\eqref{eq:Sigma_large_D}, one can derive analytically a covariance matrix for continuous gains in the limit where $\sigma^4\sim K$ (see Eqs.~\eqref{aeq:Sigma_cont_large_K} and \eqref{aeq:Sigma_cont_large_K_c=24}). The resulting structure is exactly the same as for binary gains, just the numerical prefactors differ.

Altogether, a detailed comparison of the maximum entropy distributions for binary and continuous gains shows a highly analogous structure.

\section{Comparison to networks trained with gradient descent.}

\begin{figure*}[h!]
\centering
\begin{overpic}[width=0.49\linewidth]{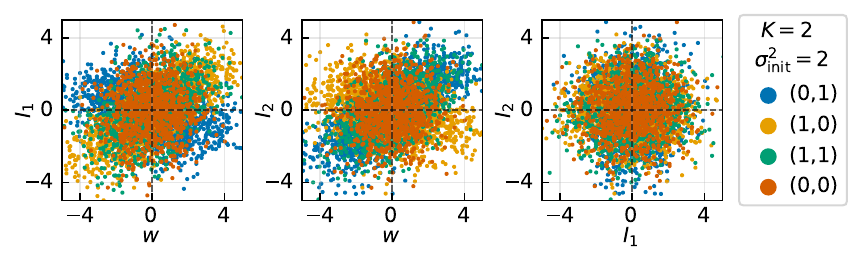}
    \put(0,27){\textbf{A}}
\end{overpic}
\hfill
\begin{overpic}[width=0.49\linewidth]{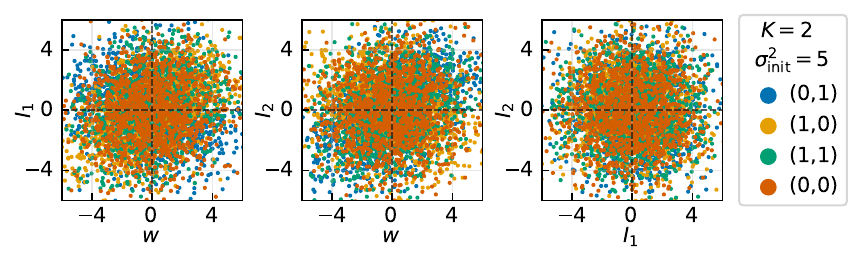}
    \put(0,27){\textbf{B}}
\end{overpic}

\begin{overpic}[height = 3.5cm]{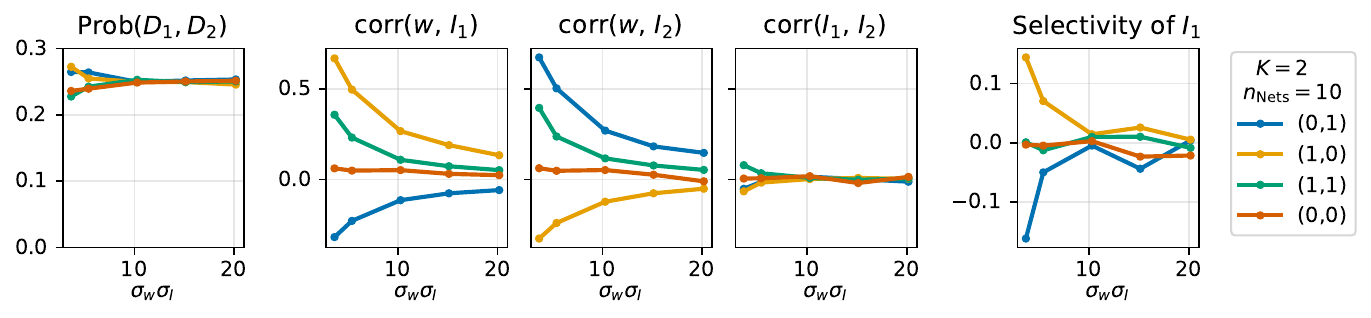}
    \put(0,21){\textbf{C}}
    \put(21,21){\textbf{D}}
    \put(70,21){\textbf{E}}
\end{overpic}

\begin{overpic}[height = 3.5cm]{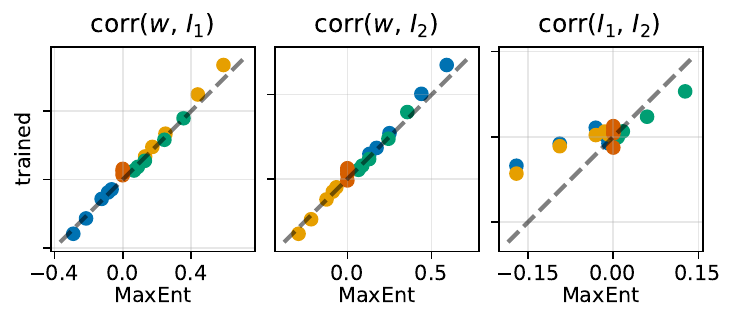}
    \put(0,40){\textbf{F}}
\end{overpic}
\begin{overpic}[height = 3.5cm]{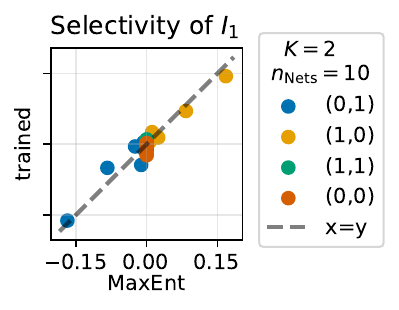}
    \put(0,73){\textbf{G}}
\end{overpic}

\begin{overpic}[height = 3.5cm]{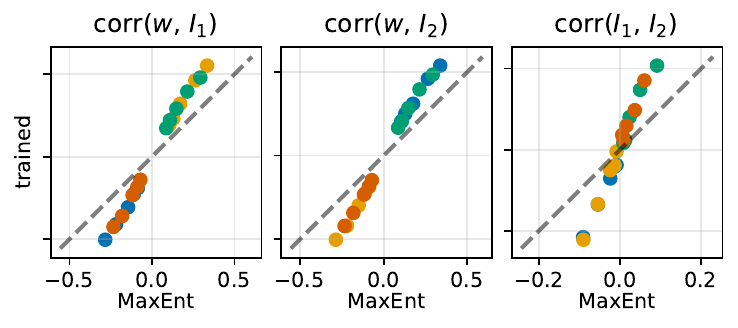}
    \put(0,40){\textbf{H}}
\end{overpic}
\begin{overpic}[height = 3.5cm]{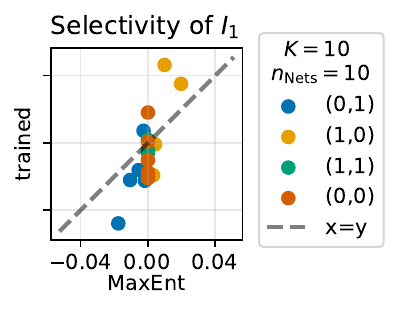}
    \put(0,73){\textbf{I}}
\end{overpic}
\caption{Connectivity structure in networks trained with gradient descent. \textbf{A, B}: Scatter plots of weights $(w_i,I_{i1},I_{i2})$ after training a ReLU non-linear network with $N=5000$ neurons on the task. The colors correspond to the  values of gains $(D_{i1},D_{i2})$ computed by linearizing the network after training (i.e. $D_{ic}=\Theta(H_{ic})$ with $\Theta$ the Heaviside function). The weights were initialized as centered Gaussians with variances  $\sigma_\mathrm{init}^2=2$ (A) and $\sigma_\mathrm{init}^2=5$ (B). \textbf{C, E, F}: Summary statistics (probability, correlation, selectivity) as function of weight variances after training, more precisely against the product  $\sigma_w \sigma_I$ of the standard deviations of output and input weights after training.  All quantities were averaged over $10$ identically trained networks.
\textbf{F-I}:  direct comparison of summary statistics (correlation and selectivity) between trained networks and  maximum entropy networks. We first train networks for a range of initial variances $\sigma_\mathrm{init}^2$, and then compute the maximum entropy distribution that corresponds to the variances of $w$ and $I$ after training, by fixing the weight scale to be $\sigma^2=\sigma_w \sigma_I$. For each network, we divide neurons into  four populations based on gains, and then plot the quantity (correlation or selectivity) corresponding to the maximum entropy distribution on the x-axis and the quantity corresponding to the trained network on the y-axis. Each point of the same color therefore corresponds to a population within a network with a different scale $\sigma^2$. \textbf{F-G}: $K=2$ contexts. \textbf{H-I}: $K=10$ contexts.
}
\label{fig:grad_desc}
\end{figure*}

We next asked how the maximum-entropy distributions compare to those obtained from the more standard approach of
adjusting parameters with gradient descent. We therefore
 trained the full non-linear network (Eq.~\eqref{eq:nonlin_ntwk}) on the context-dependent input-selection task (Eq.~\eqref{eq:task}), and examined the distribution of  parameters generated by this process. We systematically varied the scale of weights at initialization, a standard approach for controlling the learning regime of gradient descent \cite{Jacot2018-ex,Chizat2019-tt,Arora2019-kf,Lee2019-dq,Woodworth2020-gf, Geiger2020-wf,Woodworth2020-gf,Paccolata2020-em}.
 
More specifically, we used
$\phi=\mathrm{ReLU}$ as a non-linearity, and homogeneously sampled input-output pairs $\{(u;c),u_c\}$ as training data. The training algorithm was full batch gradient descent with i.i.d.\ Gaussian initialization of weights of variance $\sigma_\mathrm{init}^2$, and without weight regularization or additive white noise. We trained the parameters $w \in \mathbb R^N$ and $I,H\in \mathbb R^{N\times K}$ of the non-linear network, and  computed after training the corresponding gains $D_{ic}=\phi'(H_{ic})$, which in this case are binary.

We found that the resulting  empirical distributions of single-neuron parameters $(w_i,I_i,D_i)\in\mathbb R^{2K+1}$ bore a close resemblance
to the maximum-entropy distributions, both for small and large $\sigma_\mathrm{init}$ (compare Fig.~\ref{fig:grad_desc} A,B with Fig.~\ref{fig:binary_K=2} A,B). We then computed the same summary statistics as for the maximum-entropy distribution, and examined their values as function of weight scale for different values $K$ of the number of contexts.
More specifically, for $K=2$ and $K=10$, we grouped neurons in four populations based on  the four possible combinations of $(D_1,D_2)$. We then  computed the fraction of neurons in each of these four conditions, the correlations of $w,I_1$ and $I_2$ and the selectivity to $I_1$. In contrast to the maximum entropy approach, the variances of $w$ and $I_a$ do not remain fixed to their initial value $\sigma_\mathrm{init}^2$, but  change in the course of training. Examining the values of the summary statistics as function of the variances after training revealed a qualitative picture highly similar to the maximum-entropy approach (compare Fig.~\ref{fig:grad_desc}~C,D,E with Fig.~\ref{fig:binary_K=2}~E,F,G). 

We next quantitatively compared  the values of summary statistics across trained and maximum entropy distributions at a fixed value of variances of $w$ and $I_a$. For $K=2$ contexts, the match is almost perfect (Fig.~\ref{fig:grad_desc}~F,G), except for the correlation between $I_1$ and $I_2$. For $K=10$ contexts, the weight correlations in the trained networks are systematically larger in absolute value than corresponding quantities in the maximum entropy distribution, but remain proportional (Fig.~\ref{fig:grad_desc} H,I). 

Altogether, our results show that the maximum entropy distributions capture surprisingly well the connectivity structure in networks trained with gradient descent.
This match is not just qualitative, but even quantiative, across a large range of our two hyperparameters, the weight scale and the number of contexts. More work will be required to fully explain such a close quantitative match.


\section{Discussion}
In this work, we introduced a normative framework for determining the minimal connectivity structure that a network must possess in order to perform a given task. Rather than relying on gradient descent optimization, we characterized network connectivity as a probability distribution over single-neuron weights, and derived constraints on this distribution ensuring both compatibility with the task and appropriate scaling of the weights. Applying the maximum entropy principle then yields a unique distribution that satisfies these constraints while remaining as random as possible. This allows us to capture the structure  needed by the task, free of any bias introduced by a particular training procedure. We applied this framework on solving a context-dependent input selection task in a feed-forward network with one hidden layer, for which the maximum entropy distribution can be derived analytically by mapping the network onto a gain-modulated linear model and exploiting the symmetries of the task.

The resulting distribution has a non-trivial yet highly interpretable structure. Our central result is that task constraints induce the emergence of  neuronal populations defined by the joint statistics of contextual modulation and synaptic weights. The strength and  organization of this structure depend on the overall weight scale $\sigma^2$ and number of contexts $K$. More specifically, the maximum entropy distribution
is a mixture of Gaussians, where each population is defined by its pattern of gain values across contexts.
For $K=2$ stimuli and contexts, the network consists of four populations. Neurons in two of these populations are task specialized and each activated exclusively  in only one context. The two other populations, of smaller size, are not strictly necessary for the task, but appear because of entropy maximization. For a small weight scale $\sigma^2$, the two specialized populations show preferential selectivity to the stimulus relevant in each context, while the task irrelevant populations show random selectivity. 
With increasing weight scale $\sigma^2$, the preferential selectivity fades away and all neurons become randomly selective to any stimulus. Task performance is however ensured by keeping appropriately tuned correlations between input and output weights in the task relevant population. 
For a large number $K$ of stimuli and contexts, the organization is different as individual neurons do not specialize for individual contexts. Instead,
a random half of neurons are active in each context, and all  populations have the same size.  
This is different from a naive guess,  where each neuron is active in only one context. 
Furthermore, for small weight scale $\sigma^2$ we find a structured selectivity of input weights, but without preference for  a single stimulus. Altogether, our results show that maximizing the entropy of network parameter distribution accounts for several types of connectivity structure when varying the two  hyperparameters.

A key question is what is the biological meaning of a maximum entropy principle. For a distribution of several variables, entropy is maximal if all variables are independent. In absence of task constraints, maximizing the entropy of the weight distribution therefore leads to synapses that are random and independent of each other.
Task constraints instead induce correlations between the different synapses, specifically between  incoming and outgoing weights for each individual neuron. At the level of underlying biophysical mechanism, it is natural to assume that any coordination between synapses  requires additional metabolic costs with respect to synapses that are random and independent of each other. The maximum entropy approach provides a normative principle for balancing the randomness of independent synapses with structure induced by task constraints, irrespective of the details of biophysical mechanism that might be implementing the coordination among synapses.
This argument is related to the idea of efficient coding \cite{Barlow1961-bj,Field1994-qz,Atick1990-db,chalk_UnifiedTheory_2018}, but it focuses on the information content of the synapses, instead of the information content of the neural activity.

In biological networks, the coordination between synapses imposed by task constraints however needs to be implemented by some type of plasticity. Interestingly, in the case of context-dependent input-selection studied here,
the task constraints take the form of third-order correlations $\mathbb E[w I_a D_c]$ which are reminiscent of three-factor learning rules, where the contextual gain plays the role of an eligibility trace \cite{Fremaux2015-wt,Magee2020-rr}. Moreover, writing down the gradient-decent updates of input weights $I$ of our gain-modulated linear network, and replacing the readout by a random vector similarly to feedback alignment \cite{lillicrap_RandomSynaptic_2016} or direct feedback alignment \cite{nokland_DirectFeedback_2016}, one can reinterpret the arising terms as a gain-modulated Hebbian learning rule \cite{Fremaux2015-wt}.
Comparing more directly the outcome of  such plasticity rules with maximum-entropy connectivity is interesting direction for future research.

The different types of structure obtained when varying the weight scale $\sigma^2$ bear a close similarity with the different types of networks resulting from different regimes of gradient descent \cite{flesch_OrthogonalRepresentations_2022}. Theoretical works in machine learning have shown that initializing networks with large output weights leads to the so-called {\em kernel} or {\em lazy learning} regime, where mainly output weights are adjusted, while input weights remain close to random \cite{Jacot2018-ex,Chizat2019-tt,Arora2019-kf,Lee2019-dq,Woodworth2020-gf}. Small initialization weights instead lead to {\em feature learning} or the {\em rich regime}, where input weights align to the features of the task \cite{Saxe2019-vf,Geiger2020-wf,Woodworth2020-gf,Paccolata2020-em}. It has been argued  \cite{flesch_OrthogonalRepresentations_2022} that for the context-dependent input-selection task, rich learning leads to networks where neurons develop structured, preferential selectivity to stimuli \cite{Cohen1990-eh}, while lazy learning leads to random mixed selectivity \cite{mante_ContextdependentComputation_2013,pagan_IndividualVariability_2025}. 
Interestingly, it has been recently shown that the rich and lazy learning regimes  can also appear in more biologically plausible learning rules such as feedback-alignment or direct-feedback-alignment, depending on how the output of the network scales with the number of neurons \cite{bordelon_InfluenceLearning_2022}. In maximum-entropy networks, varying weight scales directly interpolates between different types of solutions, providing a potential normative account for the different types of network structure independently of the specific learning algorithm.


The match we found between maximum-entropy networks and gradient descent is however not only qualitative, but also quantitative. More specifically,
we found that the second order statistics of trained weights quantitatively agree with the maximum entropy distributions. 
Several theoretical studies have sought to formulate stochastic gradient descent as stochastic dynamics, which, under specific conditions, converge at equilibrium to a distribution minimizing a combination of the (negative) entropy and the loss \cite{Chaudhari2017-ml,mei_MeanField_2018, Zhang2018-wi,Adhikari2023-bi}. The properties of the stochastic dynamics and the resulting equilibrium distribution however depend on the specific assumptions for  the noise in stochastic gradient descent. Here we used noiseless, full-batch gradient-descent with a large learning rate, and it remains to be understood if and how a description in terms of stochastic dynamics is applicable. One notable difference with the maximum entropy approach is that in gradient-descent training, one cannot fix the final scales of the weights a priori. We therefore expect that gradient-descent will be closer to Bayesian inference of a posterior distribution from a Gaussian prior and a likelihood that is constructed from the loss \cite{mandt_StochasticGradient_2018}. Understanding the exact connection of maximum entropy, gradient-descent and Bayesian inference however, needs additional work.

Our analysis of maximum-entropy connectivity for context-dependent input selection relies on the fact that this task is linear within each context. This allowed us to linearize non-linear networks in each context, and  map them onto gain-modulated linear models, which are related to gated and piecewise linear models \cite{Linderman2017-hh,Saxe2022-li,Morrison2024-ed}. This analysis shows that gain patterns play a key role in defining the resulting populations and implementing the computation. We therefore treat these gains as abstract computational quantities, in the sense that their only computationally relevant  property is their three-point correlation with input and output weights. In particular, the computation and the resulting connectivity structure are independent of the original non-linearity, and  on whether the gains are binary or continuous. Ultimately, the gain-modulation needed for the task could be implemented by a variety of biological mechanisms \cite{Ferguson2020-zr}. Our results therefore provide an additional perspective on the principles of computation through gain-modulation \cite{Salinas2000-vu,Stroud2018-ol,Costacurta2024-hf}. We expect that our approach can be extended to a variety of tasks that can be approximated as linear pieces \cite{Logiaco2021-fx,Kao2021-dw}. For example, it generalizes in a straightforward way to tasks in which the stimuli have to be not only selected, but linearly combined differently in each context. The only difference to the task we have considered here, is that one might lack enough symmetry to reduce the set of Lagrange multipliers to a level that is analytically tractable.

Our approach is based  on the mean-field assumption that the distribution of parameters in the network factorizes across individual neurons, so that we   only characterize the distribution of single-neuron parameters $\theta_i = (w_i,I_i,D_i)$. Gradient-descent training in two-layer networks usually leads to such smooth factorized distributions of single-neuron parameters when the number of neurons in the hidden layer is large \cite{mei_MeanField_2018,rotskoff_TrainabilityAccuracy_2022,sirignano_MeanField_2019}. This is the original motivation for our mean-field assumption. One could ask what would happen if instead, we had allowed correlations of weights across neurons. In Appendix \ref{app:iid_neurons} we show that, when we require that the average network performs the task,
the maximum entropy approach naturally leads to a factorized a distribution of network parameters $p(\theta_1,\cdots,\theta_N)=\prod_i p(\theta_i)$. We can therefore restrict ourselves to factorized distributions without loss of generality.

In this article we have exclusively focused on context-dependent input selection without any time dependence. A straightforward generalization of our task is to allow noisy stimuli that vary in time so that temporal integration is necessary to determine a quantity of interest, for example the temporal mean \cite{mante_ContextdependentComputation_2013,pagan_IndividualVariability_2025}. Previous studies have examined the structure of recurrent neural network (RNN) models trained on this task with gradient descent \cite{mante_ContextdependentComputation_2013,Yang2019-wi,pagan_IndividualVariability_2025,Langdon2025-cv}. In particular, works with low-rank RNNs \cite{mastrogiuseppe_LinkingConnectivity_2018} have argued that performing context-dependent integration requires neurons to be organized in several populations based on their gains \cite{dubreuil_RolePopulation_2022,Valente2022-vm,Barbosa2023-di}. These analyses relied on a mean-field approach directly analogous to the one employed here for feed-forward networks, and assumed that the distribution of single-neuron parameters follows a mixture of Gaussians \cite{Beiran2021-gv,dubreuil_RolePopulation_2022,Barbosa2023-di}. This approach can be justified a posteriori by the maximum entropy approach developed here, which can be directly applied to unit-rank RNNs. Extending the maximum entropy framework to more general RNNs  is an exciting direction to be explored further.

\begin{acknowledgments}
This work was supported by a grant from the Simons Foundation (AN-NC-GB-Culmination-00003154-05, SO) and the program “Ecoles Universitaires de Recherche” launched by the French Government and implemented by the ANR, with the reference ANR-17-EURE-0017.
\end{acknowledgments}

\printbibliography[heading=bibintoc]

\clearpage
\appendix

\section{Maximum Entropy calculation}\label{app:maximum_entropy_calculation}
\subsection{Recap on Convex optimization}\label{subapp:recap_convex_optimization}
We start with a general summary of the convex optimization approach that we use to determine the maximum entropy distribution. We follow the lecture notes \cite{rosenberg_LagrangianDuality_2019}, for more details see \cite{boyd_ConvexOptimization_2023}. 

Our aim is to find a normalized probability distribution $p(\theta)$
on some domain $\theta\in I$ which maximizes the entropy $H(p)=-\int_{I}p\log p$
under constraints $\mathbb{E}[f_{k}(\theta)]=c_{k}$. This is a convex
optimization problem because:
\begin{itemize}
\item The entropy $H(p)=-\sum_{\theta}p_{\theta}\log p_{\theta}$ is strictly concave (jointly
in all $p_{\theta})$, because the function $-x\log(x)$ is concave.
\item The feasible set is convex. Focusing on a discrete distribution $p=(p_{\theta_1},\cdots,p_{\theta_n})$ of concreteness,  $\mathcal{P}=\{p\in\mathbb{R}^{n}|\,p_{\theta}\ge0,\sum_{\theta}p_{\theta}=1,\sum_{\theta}p_{\theta}f_{k}(\theta)=c_{k}\}$
is a convex set.
\end{itemize}
To maximize entropy under the constraints, the usual approach is to find the minima of the following Lagrangian, where constraints are enforced through so-called Lagrange-multipliers $\lambda$,
\begin{equation}
\mathcal{L}(p,\lambda)=\int_{I}p\log p-\sum_{k}\lambda_{k}(\int_{I}f_{k}p-c_{k}).
\end{equation}
The supremum over the Lagrange multipliers $\lambda$ is
\begin{equation}
\sup_{\lambda}\mathcal{L}(p,\lambda)=\begin{cases}
\int_{I}p\log p & \text{if for all \ensuremath{k}: }\mathbb{E}[f_{k}(\theta
)]=c_{k}\\
\infty & \text{otherwise}
\end{cases}
\end{equation}
and therefore minimization of $\sup_\lambda \mathcal L(p,\lambda)$ leads to the desired distribution $p^{*}$, that maximizes entropy under the constraints. The optimal value of the Lagrangian corresponding to $p^{*}$ is
\begin{equation}
l^{*}=\inf_{p}\sup_{\lambda}\mathcal{L}(p,\lambda).
\end{equation}
This is the primal problem. The dual problem is obtained by swapping
supremum and infimum
\begin{equation}
d^{*}=\sup_{\lambda}\inf_{p}\mathcal{L}(p,\lambda).
\end{equation}
The advantage of this is that the dual function 
\begin{align} \nonumber
g(\lambda):=&\inf_{p}\mathcal{L}(p,\lambda) \\
=&\inf_{p}\left(\int_{I}p\log p-\sum_{k}\lambda_{k}(\int_{I}f_{k}p-c_{k})\right)
\end{align}
is always concave, because it is the point-wise minimum of affine
functions (in $\lambda$). Indeed, for any $\lambda_{1},\lambda_{2}$
and $t\in[0,1]$ we have 
\begin{align} \nonumber
g(t\lambda_{1}+(1-t)\lambda_{2}) & =\inf_{p}\left[t\mathcal{L}(p,\lambda_{1})+(1-t)\mathcal{L}(p,\lambda_{2})\right]\\ \nonumber
 & \ge t\inf_{p}\mathcal{L}(p,\lambda_{1})+(1-t)\inf_{p}\mathcal{L}(p,\lambda_{2})\\
 & =tg(\lambda_{1})+(1-t)g(\lambda_{2}),
\end{align}
where in the first equal sign we used affinity of $\mathcal{L}$ in
$\lambda$. The principle of weak duality now tells us that in any
case $l^{*}\ge d^{*}$, i.e. the dual problem is a lower bound on
the optimal value. If the primal problem is convex and the optimal
point $p^{*}$ is feasible (i.e. the distribution $p^{*}$ for which
the Lagrangian takes its optimal value $l^{*}$ is included in the
feasible set $\mathcal{P}$ of points that satisfy the constraints),
then one has equality $l^{*}=d^{*}.$

This equality holds in our case of entropy maximization. The infimum
of $\mathcal{L}$ over all distributions is found by setting $\delta\mathcal{L}(p,\lambda)/\delta p(x)=0$,
and one finds the optimal distribution to be
\begin{equation} \label{eq:max_ent_general}
p_{\lambda}(x)=\frac{1}{Z(\lambda)}e^{\sum_{k}\lambda_{k}f_{k}(x)}
\end{equation}
with $Z(\lambda)=\int_{I}\exp\left(\sum_{k}\lambda_{k}f_{k}(x)\right)$.
Substituting $p_{\lambda}$ into $\mathcal{L}(p,\lambda)$, one obtains
the concave dual function
\begin{equation}
g(\lambda)=-\log Z(\lambda)+\sum_{k}\lambda_{k}c_{k}.
\end{equation}
The Lagrange multiplies $\lambda$ are found by maximizing
this function.

\subsection{Application to our setting}
We next apply this convex optimization approach to our setting. 
Our goal is to determine the distribution of the weights 
\begin{equation}
    \theta=(w,I,D)\in (\mathbb R, \mathbb R^{K},\mathcal D)
\end{equation}
with $I=(I_{1},\cdots,I_{K})$ and $D=(D_{1},\cdots,D_K)$ that maximizes the Shannon entropy
\begin{equation}
H(p):=-\int d\theta \,p(\theta)\log(p(\theta)),
\end{equation}
while satisfying the task constraints (Eq.~\eqref{eq:task-constraint}) and the scale constraints (Eq.~\eqref{eq:scale_constraints})
\begin{align} \label{eq:constraints_appendix}
    \mathbb{E}[wD_{a}I_{c}]&=\delta_{ac}, &
    \mathbb{E}[I_{a}^2]&=\sigma_{I}^{2}, &
    \mathbb{E}[w^2]&=\sigma_{w}^{2}.
\end{align}
Note that here we allow the variances of input and output weights to be different.
The domain $\mathcal D$ of the gains $D=\phi'(H)$ depends on activation function and can be either binary $\mathcal D=\{0,1\}^{K}$ (for $\phi=\mathrm{ReLU}$) or continuous
$\mathcal D=[0,1]^{K}$ (for $\phi=\mathrm{erf}$ or $\tanh$).

\subsubsection{Dual problem.}
The Lagrangian to be minimized is

\begin{align}
\mathcal{L}(p,\lambda)=&\int p\log p-\sum_{ab}\gamma_{ab}(\int wI_{a}D_{b}p-\delta_{ab}) \\ \nonumber
&+\sum_{a}\beta_{a}(\int I_{a}^{2}p-\sigma_{I}^{2})+\alpha(\int w^{2}p-\sigma_{w}^{2}),
\end{align}
that is, our Lagrange multipliers are 
\begin{equation}
    \lambda=\left(\alpha, \{\beta_a\}_{a=1}^K, \{\gamma_{ab}\}_{ab=1}^K\right).
\end{equation}
From Eq.~\eqref{eq:max_ent_general}, taking into account the form of the constraints in Eq.~\eqref{eq:constraints_appendix},
the resulting (normalized) probability
distribution is 
\begin{equation}\label{aeq:full-distribution-abgamma}
p_{\lambda}(\theta)=\frac{1}{Z_\lambda}\exp(-\alpha w^{2}-I^{T}BI+wI^{T}\gamma D),
\end{equation}
where we introduced $B=\mathrm{diag}(\beta_{1},\cdots,\beta_{K})$. The normalization constant (also called \textit{partition function}) is
\begin{equation}
Z_\lambda=\int_{\mathcal D}dD\int_{\mathbb{R}^{K}}dI\int_{\mathbb{R}}dw\,\exp(-\alpha w^{2}-I^{T}BI+wI^{T}\gamma D).
\end{equation}
We need $\alpha, \beta_a >0$
for the distribution to be normalizable. 
The corresponding concave dual problem is the maximization of $g(\lambda):=\mathcal{L}(p_{\lambda},\lambda)$, which becomes
\begin{equation}\label{aeq:dual_g_lambda}
     g(\lambda)=-\log Z_\lambda+\mathrm{Tr}(\gamma)-\sigma_{I}^{2}\mathrm{Tr}(B)-\sigma_{w}^{2}\alpha.
\end{equation}

\subsubsection{Symmetry.}
Taking into account the symmetry of the dual function  allows us to reduce
the number of independent Lagrange multipliers considerably. The dual function $g(\lambda)$ is invariant under permutations $\sigma\in S_K$ that transform on the Lagrange multipliers according to
\begin{equation}
\gamma\to P_\sigma^{T}\gamma P_\sigma,\qquad B\to P_\sigma^{T}B P_\sigma.
\end{equation}
where $P_\sigma$ is a matrix the permutes the coordinates of $\mathbb{R}^{K}$ (the so-called natural representation of $S_K$). Indeed, this transformation leaves $\mathrm{Tr}(\gamma)$ and $\mathrm{Tr}(B)$ invariant. And $Z_\lambda$ is invariant,
because we can do the variable change $I\to P_\sigma I$ and $D\to P_\sigma D$ (with Jacobi
determinant one) which also leaves the integration region $D\in [0,1]^{K}$ and $I\in \mathbb R^{K}$ invariant. This symmetry tells us that whenever $\lambda = (\alpha, B, \gamma)$ is a solution (a maximum of $g$), the transformed Lagrange-multipliers $\lambda_{\sigma}=(\alpha, P_{\sigma}^{T}BP_{\sigma}, P_{\sigma}^{T}\gamma P_{\sigma})$ must also be a solution, for all permutations $\sigma\in S_{K}$. Since the function $g$ is convex, it cannot have several isolated maxima $\lambda_\sigma$. Therefore, $\lambda=\lambda_\sigma$. This implies (by Schur's lemma) that all $\beta_{a}$ are equal and that $\gamma$ is parametrized
by only two parameters which we call $t$ and $s$ (for ``trivial''
and ``standard'' representation of $S_{K}$),
\begin{align}\label{aeq:beta_gamma_sol}
B & =\beta\mathbbm{1}\\ \nonumber
\gamma & =s\mathbbm{1}+(t-s)uu^{T}
\end{align}
where $u=\sqrt{\frac{1}{K}}(1,\cdots,1)$. Therefore 
\begin{equation}
g(\lambda)=-\log Z_\lambda+(K-1)s+t-K\beta\sigma_{I}^{2}-\alpha \sigma_{w}^{2}.
\end{equation}
The maximum entropy distribution then becomes,
\begin{equation}\label{aeq:full-distribution-abst}
    p_\lambda \propto \exp\Big[-\alpha w^{2} - \beta |I|^2 +  w\Big( s\, I \cdot D + \frac{t-s}{K}\sum_{ab} D_a I_b \Big) \Big].
\end{equation}
\subsubsection{Evaluation of  $Z_\lambda$.}
We start to evaluate the partition function $Z_\lambda$ and we use the symmetry constrained
forms of $B$ and $\gamma$ only in the end, such that the computation
applies for more general cases if needed.

To compute Gaussian integrals over $w$ and $I$ we use that $\int_{\mathbb{R}^{n}}\exp\left(-(x-b)^{T}A(x-b)\right)dx=\sqrt{\pi^{n}/\det(A)}$.
We first complete the square in the exponential
\begin{align} \label{aeq:complete-squares-calc}
 &-\alpha w^{2}-I^{T}BI+wI^{T}\gamma D \\ \nonumber
 &=-\alpha\left(w^{2}-\frac{1}{\alpha}wI^{T}\gamma D\right)-I^{T}BI\\ \nonumber
 &=-\alpha\left(w-\frac{1}{2\alpha}I^{T}\gamma D\right)^{2}-I^{T}\left(B-yy^T\right) I
\end{align}
where we introduced $y:=\gamma D/\sqrt{4\alpha}$. 
Therefore, performing integration over $w$ and $I$ yields
\begin{equation}
Z_\lambda[\mu]=\sqrt{\frac{\pi}{\alpha}}\sqrt{\pi}^{K}\int_{\mathcal D}dD\det(B-yy^{T})^{-1/2}.
\end{equation}
To further simplify, we use the
rank-1 identity
\begin{align*}
\det(B-yy^{T}) & =\det(B)(1-y^{T}B^{-1}y).
\end{align*}
Setting $B=\beta\mathbb{I}$ from the symmetry argument, we get
\begin{equation}\label{aeq:Z-mu}
Z_\lambda=\#\int_{\mathcal D}dD\,\left(1-\frac{D^{T}\gamma^{2}D}{4\alpha\beta} \right)^{-1/2}
\end{equation}
with prefactor $\# = \sqrt{\frac{\pi}{\alpha}}\sqrt{\frac{\pi}{\beta}}^{K}$. The ansatz for $\gamma$ (Eq.~\eqref{aeq:beta_gamma_sol}) can be used to write 
\begin{equation}
\gamma^{2}=s^{2}\mathbbm{1}+(t^{2}-s^{2})uu^{T}
\end{equation}
and therefore 
\begin{align} 
D^T\gamma^{2}D=s^{2}\sum_{a}D_{a}^{2}+\frac{t^{2}-s^{2}}{K}(\sum_{a}D_{a})^{2}.
\end{align}
Then the partition function becomes
\begin{equation}\label{aeq:Z-Q}
    Z_\lambda= \#\int_{\mathcal D}dD\,(1-Q(D))^{-1/2},
\end{equation}
with 
\begin{equation} \label{aeq:Q-def}
Q(D):=\frac{1}{4\alpha \beta} \left(s^{2}\sum_{a}D_{a}^{2}+\frac{t^{2}-s^{2}}{K}(\sum_{a}D_{a})^{2}\right).
\end{equation}

\subsubsection{Solving for Lagrange multipliers.}

To determine the values of the Lagrange multipliers $\lambda = (\alpha,\beta,s,t)$, we compute the derivatives $\partial_\lambda g$ and set them to zero: 
\begin{align}\label{aeq:abst_eq}
    \partial_\alpha g &= \frac 1{2\alpha} \Big(1+\frac \# {Z_\lambda} \int_\mathcal D dD (1-Q)^{-3/2} Q\Big) - \sigma_w^2 =0 \\ \nonumber
    \partial_\beta g &= \frac 1{2\beta} \Big(K+\frac \# {Z_\lambda} \int_\mathcal D dD (1-Q)^{-3/2} Q\Big) - K\sigma_I^2=0  \\ \nonumber
    \partial_s g &= -\frac \# {Z_\lambda} \int_\mathcal D dD (1-Q)^{-3/2} \frac{s K}{4\alpha \beta} (y-x^2) + (K-1)=0 \\ \nonumber
    \partial_t g &= -\frac \# {Z_\lambda} \int_\mathcal D dD (1-Q)^{-3/2} \frac{t K}{4\alpha \beta} x^2 + 1=0 .
\end{align}
From $\partial_\alpha g=0$ and $\partial_\beta g=0$, one directly finds a relation between $\alpha$ and $\beta$
\begin{equation}\label{aeq:sol-a}
    \alpha = \frac{K(2\beta \sigma_I^2 -1)+1}{2\sigma_w^2},
\end{equation}
without the need to do the integral over $D$. Furthermore, one by combining the two equations, recalling that $Q$ depends only on the product $\alpha \beta$, one sees that this product $\alpha \beta$ depends only on the product of variances $\sigma_w^2 \sigma_I^2$. Therefore, also $s$ and $t$ are functions of $\sigma_w^2 \sigma_I^2$ only. As a result, most properties of the distribution depend only on $\sigma_w^2 \sigma_I^2$, and not on the weight scales $\sigma_w^2$ and $\sigma_I^2$ individually. This justifies, why in the main text we simplified to the case $\sigma_w=\sigma_I=:\sigma$.

For a complete solution with arbitrary $K$, we compute the integral in $Z_\lambda$ numerically and then maximize $g(\lambda)$ gradient descent (see Appendix \ref{app:numerical_details}. When $K=2$ and the gains are binary, the four equations can be reduced  analytically to a single equation which then can be solved more efficiently (see Section \ref{subapp:K=2}). On the other hand, when
$K$ is large, the integrals can be evaluated analytically via a  saddle point approximation (see Section \ref{subsapp:large-K-limit}).   From now one we are going to drop the subindex $\lambda$ on $p_\lambda$ and $Z_\lambda$ assuming we are dealing only with the optimal Lagrange multipliers $\lambda$.

\subsection{Derivation of \texorpdfstring{$\mathbb{E}[w\phi(H_c)]=0$}{E[w phi(H\_c)]=0}.}\label{subapp:derivation-E[wphi(H)]=0}
Here we show that the first term $\mathbb E[w \phi(H_c)]$ in Eq.~\eqref{eq:linearization} is zero under the maximum entropy distribution, as assumed in the main text. A sufficient condition for this term to be zero is that the marginal of $w$ and $H$ has the following symmetry in $w$,
\begin{equation}
    p_{wH}(-w,H)=p_{wH}(w,H).
\end{equation}
This property is satisfied whenever it is satisfied by the marginal of $w$ and $D$ because $D=\phi'(H)$ imposes a correlation between $D$ and $H$ that does not involve $w$. The marginal of $w$ and $D$ is obtained by integration of Eq.~\eqref{aeq:full-distribution-abst} over $I$ similarly to the calculation in Eq.~\eqref{aeq:complete-squares-calc}. One finds
\begin{equation}\label{aeq:marginal_wD}
    p_{wD}(w,D)\propto e^{-\alpha(1-Q(D))w^2}
\end{equation}
with $Q$ from Eq.~\eqref{aeq:Q-def}. This distribution satisfies $p_{wD}(-w,D)=p_{wD}(w,D)$ and therefore $\mathbb E[w \phi(H_c)]=0$.

\subsection{Decomposition of the distribution}\label{subapp:decomposition}
To gain more intuition about the maximum entropy distribution Eq.~\eqref{aeq:full-distribution-abst}, we decompose it into a Gaussian and a non-Gaussian part by conditioning on gain parameters $D$,
\begin{equation}\label{aeq:p_wID-decomposed}
    p(w,I,D)=p_D(D)\, p_{wI|D}(w,I|D).
\end{equation}
From Eq.~\eqref{aeq:full-distribution-abst} one finds that
\begin{equation}
    (w,I)\sim \mathcal N(0, \Sigma(D))
\end{equation}
is a $K+1$ dimensional Gaussian distribution with a $D$-dependent covariance matrix $\Sigma(D)=M(D)^{-1}$ where
\begin{equation}
    M(D) =
    \begin{pmatrix}
        2\alpha & -(\gamma D)^T \\
        -\gamma D & 2\beta \mathbbm{1}_K,
    \end{pmatrix}
\end{equation}
with $\gamma D= sD+(t-s)\sum_b D_b/K$. Due to the block structure, it can be easily inverted, and the entries of $\Sigma(D)$ are
\begin{align}\label{aeq:Sigma(D)}
\Sigma(D)_{ww} &= \frac{1}{2\alpha (1-Q(D))}, \\ \nonumber
\Sigma(D)_{w I_a} &= \Sigma_{I_a w} = \frac{(\gamma D)_a}{4\alpha \beta (1-Q(D))}, \\ \nonumber
\Sigma(D)_{I_a I_b} &= \frac{1}{2\beta} \left(\delta_{ab} 
+ \frac{(\gamma D)_a(\gamma D)_b}{4\alpha \beta (1-Q(D))}\right).
\end{align}
Here $Q(D)=D^T\gamma^2D/4 \alpha \beta$  (Eq.~\eqref{aeq:Q-def}) pops up quite naturally. The marginal of $D$, can be inferred from Eq.~\eqref{aeq:Z-Q} to be
\begin{equation}\label{aeq:p_D}
    p_D(D) \propto \frac{1}{\sqrt{1-Q(D)}}.
\end{equation}

\paragraph*{Binary gains.}
If $D\in\{0,1\}^K$, one can further simplify the distribution of $D$. The distribution depends only on $n=\sum_a D_a$ and we have 
\begin{equation}
    p_n:=p_D(D) \propto \frac{1}{Z_D} \frac{1}{\sqrt{1-Q_n}}
\end{equation}
with
\begin{equation}
    Q_n = \frac{1}{4\alpha\beta}\left(s^{2}n+\frac{t^{2}-s^{2}}{K}n^{2}\right),
\end{equation}
and the normalization constant
\begin{equation}
    Z_D = \sum_{n=0}^K \binom{K}{n} \frac{1}{\sqrt{1-Q_n}}.
\end{equation}

\subsection{Solution for \texorpdfstring{$K=2$}{K=2} and binary gains.} \label{subapp:K=2}
In this particular case, we can reduce the Eqs.~\eqref{aeq:abst_eq} to a single equation which can be efficiently solved. This allows us to study how the probabilities for the four different configurations depend on the free parameters of our problem, the variances $\sigma_w^2$ and $\sigma_I^2$. 

Here we provide the technical details  of
the calculation which gives us an efficient numerical implementation for finding the Lagrange multipliers for $K=2$ and binary gains. The insights obtained from the calculation are discussed in the main text.
\subsubsection{Solving for Lagrange multipliers.}
Let us outline the calculation leading to this result in a few steps:
\\ \textbullet\ 
The four equations \eqref{aeq:abst_eq} reduce to
\begin{align}
    \label{aeq:1} &(2\alpha \sigma_w^2 - 1)Z_D = \frac{2 Q_1}{(1-\,Q_1)^{3/2}} + \frac{Q_2}{(1-Q_2)^{3/2}}\\ 
    \label{aeq:2} &\alpha = \frac{2\beta \sigma_I^2 -1/2}{\sigma_w^2} \\
    \label{aeq:3} &Z_D = (1-Q_1)^{-3/2}\frac{s}{4\alpha \beta} \\ 
    \label{aeq:4} &Z_D = \left( (1-Q_1)^{-3/2} + 2(1-Q_2)^{-3/2}) \right) \frac{t}{4\alpha \beta}.
\end{align}
where
\begin{align}\label{aeq:Q_0-Q_1-Q_2-Z_D}
    Q_0 = 0 \qquad 
    Q_1 = \frac{s^2 + t^2}{8 \alpha \beta} \qquad 
    Q_2 = \frac{t^2}{2\alpha\beta} \\ \nonumber
    Z_D = 1 + \frac{2}{\sqrt{1-Q_1}} + \frac{1}{\sqrt{1-Q_2}}.
\end{align}
\\ \textbullet\ 
Adding $s$ times \eqref{aeq:3} to $t$ times \eqref{aeq:4}, we get the rhs of the \eqref{aeq:1}. Therefore $Z_D$ cancels on both sides, and 
\begin{equation}\label{alpha(r)}
    \alpha = \frac{s+t+1}{2\sigma_w^2}.
\end{equation}
Furthermore, substituting this into \eqref{aeq:2} we have
\begin{equation}\label{beta(r)}
    \beta = \frac{s+t+2}{4\sigma_I^2}.
\end{equation}
\\ \textbullet\ 
We now introduce the ratio $r = s/t$ with the aim to express all quantities as functions of $r$ so that  are left with a single equation. Subtracting \eqref{aeq:3} from \eqref{aeq:4} one finds
\begin{equation}\label{aeq:G(r)-def}
    G(r):=\left(\frac{r-1}{2}\right)^{2/3} = \frac{1-Q_1}{1-Q_2}.
\end{equation}
Furthermore, from the definition of $Q_1$ and $Q_2$ in Eq.~\eqref{aeq:Q_0-Q_1-Q_2-Z_D}  one finds
\begin{equation} \label{aeq:Q_1(r)}
    Q_1 = \frac{r^2+1}{4}Q_2.
\end{equation}
Substituting this into Eq.~\eqref{aeq:G(r)-def}, we get
\begin{equation} \label{aeq:Q_2(r)}
    Q_2(r) = \frac{G(r)-1}{G(r)-\frac{r^2+1}{4}}.
\end{equation}
Now $Q_2$ is a function of $r$ only. Therefore, also $Q_1=Q_1(r)$ and  $Z_D=Z_D(r)$.

We do the same for the parameter $t$. From equation \eqref{aeq:3} together with the definition of $Q_2=t^2/(2\alpha \beta)$ one gets,
\begin{equation}\label{aeq:t(r)}
    t(r) = \frac{r Q_2(r)}{2Z_D(r)(1-Q_1(r))^{3/2}}.
\end{equation}
\\ \textbullet\ 
Finally, we get an equation involving $c:=\sigma_w^2 \sigma_I^2$ by equating $2\alpha \beta$ as calculated from Eq.~\eqref{alpha(r)},\eqref{beta(r)},
\begin{equation}
    2\alpha\beta = \frac{(t(r)(r+1)+1)(t(r)(r+1)+2}{4 c}
\end{equation}
to $2 \alpha \beta $ as obtained from the definition of $Q_2$. Then we have
\begin{equation}\label{aeq:f(r)}
    c=f(r):=Q_2(r) \frac{(t(r)(r+1)+1)(t(r)(r+1)+2)}{4 t(r)^2}.
\end{equation}
This is a single equation for $r$ only which can be solved easily numerically. Note that from Eq.~\eqref{aeq:Q_2(r)} we get $r>\sqrt{3}$ to ensure $Q_2<1$ and furthermore from Eq.~\eqref{aeq:G(r)-def} we get $r<3$ to be compatible with $Q_1>Q_2$ as imposed by Eq.~\eqref{aeq:Q_1(r)}.
\\ \textbullet\ 
We use the following numerical routine:
\begin{itemize}
    \item[$\circ$] Define $f(r)$ with the help of $G(r)$ Eq.~\eqref{aeq:G(r)-def}, $Q_1(r)$ Eq.~\eqref{aeq:Q_1(r)}, $Q_2(r)$ Eq.~\eqref{aeq:Q_2(r)}, $t(r)$ Eq.~\eqref{aeq:t(r)} and $Z_D(r)$ Eq.~\eqref{aeq:t(r)}.
    \item[$\circ$] For a given $c$, solve $f(r)=c$
    \item[$\circ$] For $n=0,1,2$, obtain
    \begin{equation}
        p_n(c) = \frac{1}{Z_D(r)\sqrt{1-Q_n(r)}}.
    \end{equation}
    In particular, the probabilities only depend on $c$ but not on $\sigma_w^2$ and $\sigma_I^2$ individually.
    \item[$\circ$] Obtain the parameters
    \begin{align}
        t&=t(r) & 
        \alpha &= \frac{r t(r)+t(r)+1}{2\sigma_w^2} \\
        s&= r t(r) &
        \beta&=\frac{r t(r)+t+2}{4\sigma_I^2}.
    \end{align}
    Only $\alpha$ and $\beta$ depend on $\sigma_w^2$ and $\sigma_I^2$ individually.
\end{itemize}
\textbullet\
One can furthermore observe that the function $f(r)$ is only defined for $r\in(\sqrt{3},3)$, where it is monotonically increasing. Therefore, the limiting value for $c\to\infty$ is $r=3$, which implies $Q_2=0$ from Eq.~\eqref{aeq:Q_2(r)} and $Q_1=0$ from Eq.~\eqref{aeq:Q_1(r)}. Therefore, we get $p_0=p_1=p_2=1/4$ as the asymptotic probabilities. On the other hand, one evaluates numerically that $c_{min}:= f(\sqrt{3})=1.87$, which provides a lower bound for admissible values of $c=:\sigma_w^2 \sigma_I^2$ (for $K=2$ and binary case).

\subsection{Large \texorpdfstring{$K$}{K} limit.}\label{subsapp:large-K-limit}
When $K$ is large, the integrals in Eq.~\eqref{aeq:abst_eq} determining the Lagrange multipliers can be evaluated using the central limit theorem and performing a \textit{saddle-point} approximation. To do so, note that the function $Q(D)$ from Eq.~\eqref{aeq:Q-def} has a nice structure: It only depends on the macroscopic variables
\begin{align}\label{aeq:x-y-def}
    x&:=\frac 1 K \sum_{a=1}^K D_a, & y&:=\frac 1 K \sum_{a=1}^K D_a^2.
\end{align}
We therefore rewrite
\begin{equation}\label{aeq:Q(x,y)}
    Q(D)\equiv Q(x,y) = \frac K {4\alpha \beta} \Big(s^2 y + (t^2-s^2)x^2\Big).
\end{equation}
Instead of integrating over $D$, we can now integrate over  $x$ and $y$. Since they are constructed as a sum of uniform i.i.d.\ integration variables $D_a$, according to the central limit theorem, they behave like Gaussian random variables
\begin{equation}\label{aeq:xy-distribution}
    (x,y) \sim \mathcal N \left( (\bar x, \bar y), \frac 1 K \Sigma\right)
\end{equation}
with means
\begin{align}
    \bar x&=1/2 & \bar y&=\begin{cases} 1/2, & \text{binary $D$} \\ 1/3,&\text{continuous $D$} \end{cases}.
\end{align}
Since their variances scale as $1/K$, in the large $K$ limit the distribution concentrates around the mean.
One can then approximate the integrals over their density $d\rho(x,y)$ by simply evaluating the integrand at their means (saddle-point approximation). For example the integral appearing in Eq.~\eqref{aeq:Z-Q} becomes,
\begin{align} \label{aeq:Q-saddelpoint}
    \int dD \frac 1 {\sqrt{1-Q(D)}}&\approx \int d\rho(x,y) \frac 1 {\sqrt{1-Q(x,y)}} \\  \nonumber
    &\approx \frac 1 {\sqrt{1-Q(\bar x, \bar y)}}.
\end{align}
A recurring quantity will be
\begin{equation}\label{aeq:var-D}
    \sigma_D^2 := \bar y - \bar x^2 = \begin{cases} 1/4, & \text{binary $D$} \\ 1/12, &\text{continuous $D$} \end{cases}
\end{equation}
which is the variance of a uniform integration variable $D_a\in[0,1]$ (or $D_a\in\{0,1\}$).

With this notation, the four equations \eqref{aeq:abst_eq} for the parameters $(\alpha,\beta,s,t)$ become
\begin{align}
    2\alpha \sigma_w^2 &= \frac 1 {1-Q(\bar x, \bar y)} \\
    2\beta K \sigma_I^2 &= K + \frac {Q(\bar x, \bar y)} {1-Q(\bar x, \bar y)} \\
    (K-1) \frac{4 \alpha \beta }{Ks} &= \frac{\sigma_D^2}{1-Q(\bar x, \bar y)}  \\
     \frac{4 \alpha \beta}{Kt} &=\frac{\bar x^2}{1-Q(\bar x, \bar y)}.
\end{align}

\subsubsection{Optimal parameters.}
After solving these equations  one obtains the following solutions for $(\alpha,\beta,s,t)$. We write them as a function of the parameter $t$, together with their leading order in $K$ when $\sigma_w^2\sigma_I^2 \sim \mathcal O(1)$
\begin{align} \label{aeq:abst_largeK_solutions}
    \alpha &= \frac{K(\sigma_I^2 \sigma_w^2  \bar x^2 K t  -1)+1}{2\sigma_w^2} 
    \sim \frac{K}{2 \sigma_w^2 z} 
    \\ \nonumber
    \beta &= \frac 1 2 \sigma_w^2 \bar x^2 K t 
    \sim \frac{1}{2 \sigma_I^2} \frac{z+1}{z}
    \\ \nonumber
    s &= \frac {\bar x^2}{\sigma_D^2} (K-1)t 
    \sim \frac 1 z
    \\ \nonumber
    t & =  \frac {K \sigma_D^2} { K^2 \bar x^2 z + 2 K\bar x^2 - \bar y } \sim \frac{\sigma_D^2}{\bar x^2 z K} \nonumber
\end{align}
where
\begin{equation}
    z:= \sigma_I^2 \sigma_w^2\sigma_D^2-1.
\end{equation}

\begin{figure}
    \centering
    \begin{overpic}[width=0.48\linewidth]{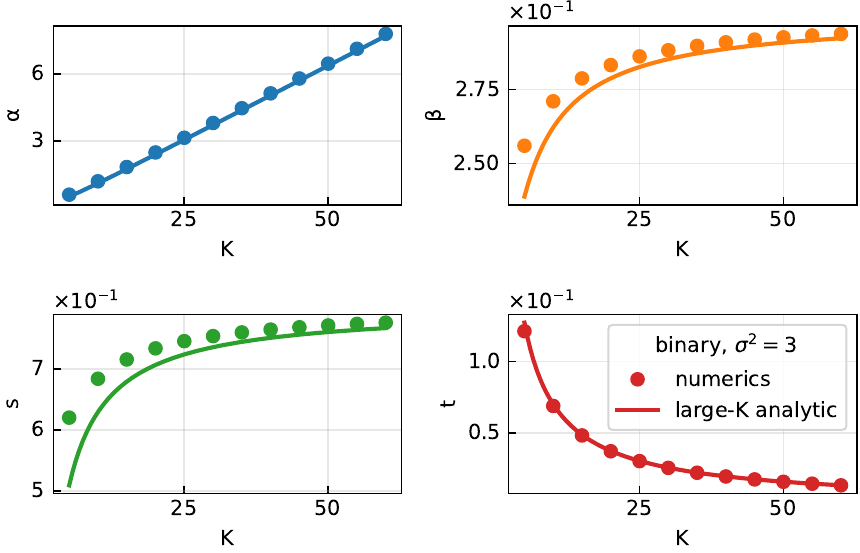}
        \put(0,62){\textbf{A}}
    \end{overpic}
    \hspace{0.2cm}
    \begin{overpic}[width=0.48\linewidth]{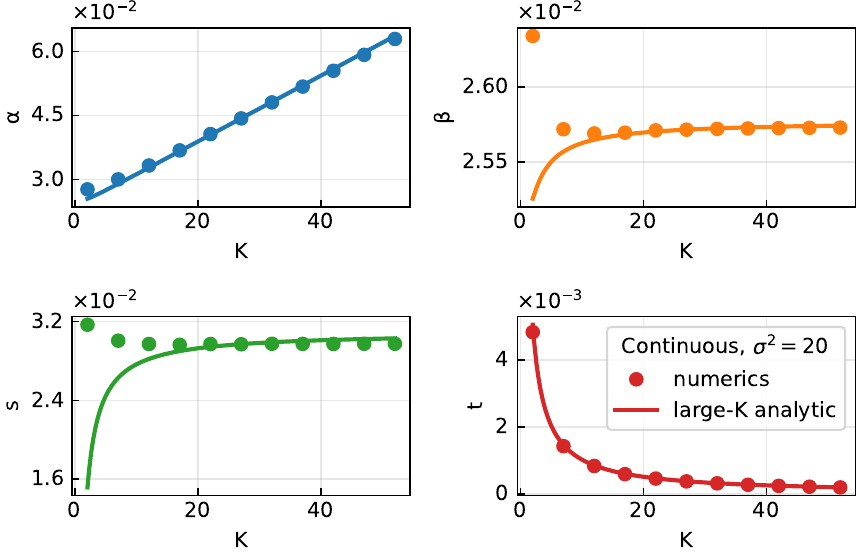}
        \put(0,62){\textbf{B}}
    \end{overpic}
    \caption{Comparison between numerical solutions (data points) and analytical large-$K$ solutions (solid lines) for $(\alpha,\beta,s,t)$. \textbf{A}: Binary $D\in\{0,1\}^K$ with $\sigma_w^2= \sigma_I^2=3$. \textbf{B}: Continuous $D\in[0,1]^K$ with larger variance $\sigma_I^2=\sigma_w^2=20$. The constraint $Q<1$ from Eq.~\eqref{aeq:constraint_c_Q<1} is only satisfied for $K<400/24\approx 17$, but the fit is also good for larger values of $K$.}
    \label{afig:numerical_params_largevar}
\end{figure}

Figure~\ref{afig:numerical_params_largevar} shows that numerical solutions of $(\alpha,\beta,s,t)$ for finite $K$ converge nicely to the analytic solutions for large $K$.

With these parameters, the distribution Eq.~\eqref{aeq:full-distribution-abst} at leading order in $K$ becomes
\begin{equation}\label{aeq:full-distribution-largeK}
    p \propto \exp\left(-\frac{Kw^{2}}{2\sigma_w^2 z} - \frac{\sum_a I_a^2}{2\sigma_I^2 \frac{z}{z+1}} + \frac{\sum_a wI_a (D_a-x)}{z}\right)
\end{equation}
where the last term $wI_a x$ couples $I_a$ to all $D_b$. 

\subsubsection{Constraints on weight variances}
\begin{figure}
    \centering
    \includegraphics[width=0.35\linewidth]{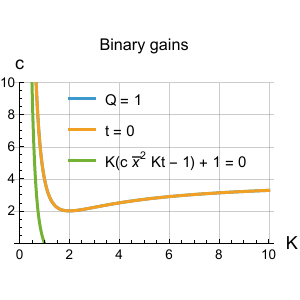}
    \hspace{1cm}
    \includegraphics[width=0.35\linewidth]{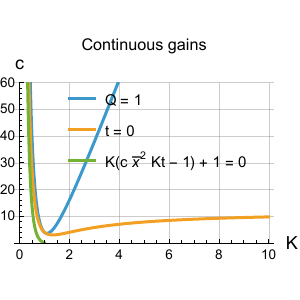}
    \caption{Phase diagram for $c:=\sigma_I^2 \sigma_w^2$ and $K$ where allowed combinations $(c,K)$, i.e. values with $\alpha,\beta>0$ and $Q<1$, lie above the three curves. For the condition $\max[Q(x,y)]<1$ we use the analytic calculation around Eq.~\eqref{aeq:V(D)} that tells us that the maximum is reached at $x=y=1/2$, but otherwise we keep the full $K$ dependence, such that the curves plotted here converge to the expression in Eq.~\eqref{aeq:constraint_c_Q<1} only for large $K$, but might differ for small $K$. For binary gains (left) the curves for $t=0$ and $Q=1$ overlap. For continuous gains (right) they are different.}
    \label{afig:range_cK}
\end{figure}

By construction, to ensure that the distribution is normalizable, we require $\alpha>0$ and $\beta>0$. This translates into two conditions for the product of the variances $c:=\sigma_I^2\sigma_w^2$,
\begin{align}
    t&>0, & K(c \bar x^2 K t -1)+1&>0.
\end{align}
More importantly, to avoid imaginary probabilities in Eq.~\eqref{aeq:p_D}, we need $\max_D[Q(D)]<1$. The three conditions are illustrated in Fig.~\ref{afig:range_cK} where the allowed values of $c$ lie above the curves. 
We now analytically simplify the last condition ($Q(D)<1$), for which we need to distinguish two regimes.

The first regime is $c\ll K $. Inserting the optimal parameters Eq.~\eqref{aeq:abst_largeK_solutions} into $Q(x,y)$ from Eq.~\eqref{aeq:Q(x,y)}, and expanding in $K$, one finds
\begin{equation}\label{aeq:Q(D)_regime1}
    Q(D)=\frac{V(D)}{\sigma_D^2} \left(1-\frac{c\sigma_D^2 - 1}{K}\right)+\mathcal O(K^{-2}).
\end{equation}
The dependence on $D$ is only through the combination 
\begin{align}\label{aeq:V(D)}
    &V(D):=y-x^2 = \frac{1}{K}\sum_a D_a^2 - (\frac{1}{K}\sum_a D_a)^2 \\ \nonumber
    &\text{with }\max_{D\in[0,1]^K} V(D) = 1/4.
\end{align}
The maximum of $V$ is reached for any $D\in\{0,1\}^{K}$ with $x=y=\frac 1 K \sum_a D_a=1/2$. One sees that for binary gains ($\sigma_D^2=1/4$) the condition $\max_D Q(D)<1$ is satisfied whenever $c>4$. For continuous gains ($\sigma_D^2=1/12$), the condition is never met, except if $c\sim \mathcal O(K)$, which leads to the following regime. 

In the second regime $c/K=:\tilde c \sim \mathcal O (1)$. Then the expansion in $K$ yields
\begin{equation}\label{aeq:Q(D)_regime2}
    Q(D) = \frac{V(D)}{\sigma_D^2 (1+ \tilde c\sigma_D^2)} + \mathcal O (K^{-1})
\end{equation}
As a sanity check, binary gains ($\sigma_D^2=1/4$) indeed satisfies $\max_D Q(D)<1$ for any value of $\tilde c$. For continuous gains ($\sigma_D^2=1/12$), however, one needs $\tilde c >24$.

Together we found that in the large $K$ limit, we need 
\begin{equation}\label{aeq:constraint_c_Q<1}
    c > \begin{cases}
        4, & \text{binary gains} \\
        24 K, & \text{continuous gains}
    \end{cases}
\end{equation}
in order to have a well-behaved probability distribution.

\subsubsection{Distribution of $D$}\label{subapp:Distribution_D_largeK}
Distinguish  the two regimes discussed in the last paragraph, we substitute the corresponding expression for $Q$, i.e.\ Eq.~\eqref{aeq:Q(D)_regime1} or Eq.~\eqref{aeq:Q(D)_regime2}, into $p_D$ from Eq.\eqref{aeq:p_D}. 

The regime $c\ll K$ only applies to binary gains for which we get
\begin{equation}\label{eq:p_D_binary_K}
    p_D (D) \propto \frac{1}{\sqrt{1 - 4V(D)(1-\lambda)}},
\end{equation}
with $\lambda:=\frac{4c - 1}{K}$. The distribution is peaked at any $D$ with $x\to 1/2$ and $\lambda$ acts as a regularizer that ensures that the distribution remains normalizable. For large $K$, we have $\lambda\to0$ and the distribution converges to a delta distribution,
\begin{equation}\label{aeq:p_D-binary-largeK}
    p_D(D)=\delta(x-1/2).
\end{equation}
This means that only those gain patterns with an equal number of zeros have non-zero probability.

The regime $c/K=\tilde c \sim \mathcal O(1)$ applies to both, binary and continuous gains. We find
\begin{equation}
    p_D (D) \propto \frac{1}{\sqrt{1 - V(D)\xi}}.
\end{equation}
with $\xi:=1/(\sigma_D^2 (1+ \tilde c\sigma_D^2))$. Whenever $\xi=4$, the distribution has a delta peak at the maximum of $V(D)$, i.e. at any \ $D\in\{0,1\}^K$ with $x=1/2$. For $\sigma_D^2=1/4$ (binary gains), we have $\xi=4/(1+\tilde c/4)\in(0,4)$. And therefore, for increasing $\tilde c$, the distribution becomes less and less peaked and uniform for large $\tilde c$. However, for $\sigma_D^2=1/12$ and $\tilde c=24$ (continuous gains), we have $\xi = 4$ and the distribution is peaked. Then for bigger values of $\tilde c$, the distribution also becomes more and more uniform. 

In summary, we have for binary gains
\begin{equation}
    p_D(D) = \begin{cases}
        \delta(x-1/2) & \text{if $\tilde c\to 0$}, \\
        \mathrm{const} & \text{if $\tilde c\to\infty$}.
    \end{cases}
\end{equation} 
and for continuous gains
\begin{equation}
    p_D(D) = \begin{cases}
        1_{D\in\{0,1\}^K}\delta(x-1/2), & \text{if $\tilde c\to 24$}, \\
        \mathrm{const} & \text{if $\tilde c\to\infty$},
    \end{cases}
\end{equation}
and for intermediate values of $\tilde c$, the distribution interpolates between the two extremes. For binary gains, the two cases are actually the same (up to $\mathcal O(K^{-1/2})$ corrections), because for uniform $D$ we have $x=1/2+\mathcal O(K^{-1/2})$. For continuous gains, however, the two cases are different. For example, in the first one we have $y=\frac 1 K \sum_a D_a^2=1/2$ and in the second one we have $y=1/3$, leading to slightly different expressions for the covariance matrix between $w$ and $I$ as we show below.

Note that the distribution $p_D$ in Eq.~\eqref{aeq:p_D} depends on the gains $D$ only through the macroscopic variables $x$ and $y$ that appear in $Q(x,y)$. Since the contribution of a single gain value $D_1$ to those macroscopic variables $x$ and $y$ is $\mathcal O(1/K)$, the variable $D_1$ becomes independent of the other gains $(D_2,\cdots,D_K)$ up to $\mathcal O(1/K)$ corrections,
\begin{equation}
    p_D \approx p_{D_1} p_{D_2\cdots D_K}
\end{equation}
where $D_1$ is uniformly distributed in $\{0,1\}$ or $[0,1]$. 

\subsubsection{Covariance structure of $w$ and $I$}
We return to the covariance matrix $\Sigma(D)$ Eq.~\eqref{aeq:Sigma(D)} to evaluate the Gaussian part of our distribution. We do this, once assuming that $p_D$ is peaked at configurations with $x=y=1/2$ and once assuming it is uniform.

\paragraph*{Binary gains.}
In the regime $c\ll K$, the distribution $p_D$ is peaked at $x=1/2$. Substituting $Q(D)$ from Eq.~\eqref{aeq:Q(D)_regime1} with $x=1/2$ together with the large $K$ expansion of $\alpha,\beta,s,t$ from Eq.~\eqref{aeq:abst_largeK_solutions} and $\sigma_D^2=1/4$, the entries of $\Sigma(D)$ are ($a\neq b$),
\begin{align} \label{aeq:Sigma_binary_large_K}
\Sigma_{ww} &= \sigma_w^2  & \Sigma_{wI_a} &= 4 (D_a-1/2) \\ \nonumber
\Sigma_{I_a I_a} &= \sigma_I^2 & \Sigma_{I_a I_b} &= \frac{16}{\sigma_w^2}(D_a-1/2)(D_b-1/2).
\end{align}
In particular, the distribution of $(w,I_a)$ or $(I_a,I_b)$ only depends on $D_a$ or $(D_a,D_b)$, but not on the other gains.
The correlation between $(w,I_a,I_b)$ is (for $a \neq b$)
\begin{align}
    \mathrm{Corr}(w,I_a|D_a) &= \frac{4(D_a-1/2)}{\sigma_w\sigma_I} \\
    \mathrm{Corr}(I_a,I_b|D_a,D_b) & = \frac{4(D_a-1/2) \, 4(D_b - 1/2)}{\sigma_w^2\sigma_I^2}.
\end{align}
and only depends on the product $c=\sigma_w^2\sigma_I^2$. The expression agrees nicely with what we find numerically for ${K=10}$ in Fig.~\ref{fig:binary_K=2}F. Furthermore, to compare to Fig.~\ref{fig:binary_K=2}G, one sees that the variance of $\mathbb E[I_a^2|D]=\sigma_I^2$ is independent of $D$ and therefore the selectivity measure we computed in this figure is zero.

In the regime $c/K=\tilde c\gg 1$, where $D$ is uniformly distributed, we have again $x=1/2$ plus fluctuations of order $1/\sqrt{K}$. Since $Q(D)$ in the binary case ($x=y$) depends on $D$ only through macroscopic variable $x$, it turns out that the covariance matrix $\Sigma(D)$ is exactly the same as before (Eq.~\eqref{aeq:Sigma_binary_large_K}), but now with variances $\sigma_w^2$ and $\sigma_I^2$ that scale with $K$. As a result the off-diagonal entries of $\Sigma$ become suppressed in $K$ compared to the diagonal entries.

\paragraph*{Continuous gains.}
Now we have to consider the regime $c/K=\tilde c \ge 24$, where $p_D$ is peaked at $x=y=1/2$ if $\tilde c\to 24$ and $p_D$ is uniform when $\tilde c \to \infty$. 

We first consider $\tilde c \to 24$. Substituting $Q(D)$ from Eq.~\eqref{aeq:Q(D)_regime2} with $x=y=1/2$, and redoing the large $K$ expansion of $\alpha,\beta,s,t$ from Eq.~\eqref{aeq:abst_largeK_solutions}, but now for ${c=24 K}$ and using $\sigma_D^2=1/12$, the entries of $\Sigma(D)$ are,
\begin{align} \label{aeq:Sigma_cont_large_K_c=24}
\Sigma_{ww} &= \frac 2 3 \sigma_w^2 K \\ \nonumber
\Sigma_{wI_a} &= 8 (D_a-1/2) K \\ \nonumber
\Sigma_{I_a I_b} &=\sigma_I^2 \Big(\delta_{ab} + 4(D_a-1/2)(D_b-1/2)\Big).
\end{align}
The structure is very similar to the binary case. Interestingly, the covariances related to $w$ scale with an additional factor $K$, such that one needs $\sigma_I\sim K$ and $\sigma_w\sim 1$ to have all entries of $\Sigma$ of same order.

In the case where $\tilde c \to \infty$, $D$ is uniformly distributed and we have $x=1/2$ and $y=1/3$ plus sub-leading fluctuations. Substituting this into $Q(D)$ from Eq.~\eqref{aeq:Q(D)_regime2} one gets
\begin{align} \label{aeq:Sigma_cont_large_K}
\Sigma_{ww} &= \sigma_w^2  \\ \nonumber
\Sigma_{wI_a} &=12(D_a-1/2) \\ \nonumber
\Sigma_{I_a I_b} &=(\sigma_I^2 - \frac{12}{\sigma_w^2})\delta_{ab} + \frac{36 \cdot 4}{\sigma_w^2}(D_a-1/2)(D_b-1/2).
\end{align}
As for binary gains, in this case, the diagonal entries of $\Sigma$ become more dominant than the off-diagonal entries, in terms of theirs scaling with $K$.

\section{Maximum entropy implies i.i.d.\ neurons}\label{app:iid_neurons}
Instead of restricting ourselves to the class mean-field network of the form Eq.~\eqref{eq:mean-field-network} in which single-neurons weights $\theta_i = (w_i,B_i)$ are independently and identically distributed with $\rho$, we could have started from a generic distribution $\tilde \rho$ of all weights $\Theta = (\theta,\cdots,\theta_N)$. We show here that in this case the maximum entropy principle leads to a factorized distribution $\tilde \rho (\Theta)=\prod_i \rho(\theta_i)$, which justifies why we directly focus on $\rho$ in the main text. 

Since the output of a network with weights sampled from $\tilde \rho$ will in general fluctuate from realization to realization, we enforce the task by requiring that the average network solves the task,
\begin{equation}
    \mathbb E[\hat f(u,e_c)]:=\frac{1}{N}\sum_i \mathbb E_i[ w_i \phi(I_i^T u+ H_i^T e_c)]\overset{!}{=}u_c.
\end{equation}
Linearizing for small $u$ then leads to the task constraints
\begin{align}
    \frac{1}{N}\sum_i \mathbb E[w_i H_{ic}]&=0, 
    &
    \frac{1}{N}\sum_i \mathbb E[ w_i\phi'(H_{ic})I_{ia}]&=\delta_{ab}.
\end{align}
These constraints do not couple different neurons together. Therefore, the maximum entropy distribution with these constraints factorizes. Note that in this case the network automatically becomes self-averaging, that is, the fluctuations of the output $\hat f(u,e_c)$ are of order $\mathcal O(N^{-1/2})$.

Let us end this section with a little comment on what happens when training such a network with stochastic gradient descent (SGD). Here the training dynamics actually couples neurons together. That is, the update of one neuron depends on the state of all the other neurons. However, when $N$ is large, then a given neuron is only affected by the \textit{mean field} of all other neurons, and its back-action on the mean field is negligible. So neurons effectively decouple and a SGD trained two-layer network with i.i.d.\ initialization is well described by a distribution over the single-neuron weights \cite{mei_MeanField_2018,rotskoff_TrainabilityAccuracy_2022,sirignano_MeanField_2019}. This is a phenomenon known in probability theory as \textit{propagation of chaos} \cite{sznitman_TopicsPropagation_1991}.

\section{Numerical details}\label{app:numerical_details}

\subsection{Solving for optimal parameters \texorpdfstring{$(\alpha, \beta, s, t)$}{(alpha, beta, s, t)}}
We solve for the optimal parameters numerically by maximizing Eq.~\eqref{aeq:dual_g_lambda} with gradient descent in pytorch. For binary $D$ we compute $Z_\lambda$ as a sum over $2^K$ terms. For continuous $D$ we do quasi-Monte Carlo integration. In both cases it is crucial to respect the constraint Eq.~\eqref{aeq:constraint_c_Q<1} on permissible $\sigma_I^2$ and $\sigma_w^2$ in order to converge to the unique maxima of the function $g(\lambda)$ avoiding regions of $\lambda$ where $Q(D)<1$.

\subsection{Sampling from the MaxEnt distribution}
We sample from the maximum entropy distribution $p$ by decomposing it into $p_{wI|D} \, p_{D} $ according to Eq.~\eqref{aeq:p_wID-decomposed}. We sample gain values from 
\begin{equation}
    p_D(D) \propto (1-Q(D))^{-1/2}
\end{equation}
 via standard Metropolis-Hastings Monte-Carlo. Here $D$ undergoes a random walk $D'=D+\xi$ with $\xi\in\mathcal N(0,\sigma^2 \mathbbm{1}_K)$. Whenever the walk goes outside the region $[0,1]^K$ it is reflected along the boundary back into the region. Acceptance of the step $D\to D'$ occurs with probability
\begin{equation}
    \alpha(D\to D')=\min(1, \frac{p_D(D')}{p(D)}).
\end{equation}
The advantage of this algorithm is that one never needs to compute the normalization of $p_D$, because one can rewrite it, such that acceptance occurs whenever
\begin{equation}
    \log u < \log p_D(D') - \log p_D(D)
\end{equation}
where in every step one draws a sample $u\sim \mathrm{Unif}(0,1)$ from the uniform distribution.

\end{document}